\documentclass[12pt]{emulateapj}
\usepackage{natbib}
\usepackage{color} 
\usepackage{gensymb} 

\newcommand{\kms}{\ifmmode {\rm km\ s}^{-1} \else km s$^{-1}$\fi}
\newcommand{\Halpha}{\ifmmode {\rm H}\alpha \else H$\alpha$\fi}
\newcommand{\Hbeta}{\ifmmode {\rm H}\beta \else H$\beta$\fi}
\newcommand{\Hgamma}{\ifmmode {\rm H}\gamma \else H$\gamma$\fi}
\newcommand{\Hdelta}{\ifmmode {\rm H}\delta \else H$\delta$\fi}
\newcommand{\Lya}{\ifmmode {\rm Ly}\alpha \else Ly$\alpha$\fi}
\newcommand{\Lyb}{\ifmmode {\rm Ly}\beta \else Ly$\beta$\fi}
\newcommand{\HeI}{\ifmmode {\rm He}\,{\sc i}\,\lambda5876 \else 
	          He\,{\sc i}\,$\lambda5876$\fi}
\newcommand{\HeII}{\ifmmode {\rm He}\,{\sc ii}\,\lambda4686 \else 
	           He\,{\sc ii}\,$\lambda4686$\fi}
\newcommand{\heii}{He\,{\sc ii}}
\newcommand{\hei}{He\,{\sc i}}

\newcommand{\feii}{Fe\,{\sc ii}}
\newcommand{\ciii}{\ifmmode {\rm C}\,{\sc iii} \else C\,{\sc iii}\fi}

\newcommand{\oiii}{O\,{\sc iii}}
\newcommand{\ob}{[O\,{\sc iii}]\,$\lambda \lambda 4959,5007$}

\newcommand{\mbh}{$M_{\rm BH}$\ }
\newcommand{\msigma}{$M_{\rm BH}$--$\sigma_{*}$\ }

\newcommand{\mrkAmbh}{$  7.25^{+  0.10}_{-  0.10}$} 
\newcommand{\mrkBmbh}{$  7.86^{+  0.20}_{-  0.17}$} 
\newcommand{\threecmbh}{$  7.84^{+  0.14}_{-  0.19}$} 
\newcommand{\pgmbh}{$  6.92^{+  0.24}_{-  0.23}$} 
\newcommand{\threecinclination}{$  17.6^{+   5.4}_{-   3.3}$} 

\newcommand{\avgfmysample}{$0.45~\pm~0.32$}
\newcommand{\dispavgfmysample}{$0.49~\pm~0.35$}

\newcommand{\avgfwholesample}{$0.54~\pm~0.17$} 
\newcommand{\dispavgfwholesample}{$0.39~\pm~0.23$} 

\newcommand{\avgffwhmmysample}{$0.45~\pm~0.33$} 
\newcommand{\dispavgffwhmmysample}{$0.52~\pm~0.36$} 

\newcommand{\avgffwhmwholesample}{$0.18~\pm~0.23$} 
\newcommand{\dispavgffwhmwholesample}{$0.59~\pm~0.22$} 

\shorttitle{Dynamical Modeling of the BLR}
\shortauthors{Grier et al.}

\begin{document}

\title{The Structure of the Broad-Line Region in Active Galactic Nuclei. II. Dynamical Modeling of Data From the AGN10 Reverberation Mapping Campaign}

\author{C.~J.~Grier\altaffilmark{1,2,3},
A.~Pancoast\altaffilmark{4,5},
A.~J.~Barth\altaffilmark{6},
M.~M.~Fausnaugh\altaffilmark{3}, 
B.~J.~Brewer\altaffilmark{7},
T.~Treu\altaffilmark{8},
\& B.~M.~Peterson\altaffilmark{3,9,10}
}
\altaffiltext{1}{Department of Astronomy and Astrophysics, Eberly
  College of Science, Penn State University, 525 Davey Laboratory,
  University Park, PA 16802; grier@psu.edu}
\altaffiltext{2}{Institute for Gravitation \& the Cosmos, The Pennsylvania State University, University Park, PA 16802, USA}
\altaffiltext{3}{Department of Astronomy, The Ohio State University, 140 W 18th Ave, Columbus, OH 43210, USA} 
\altaffiltext{4}{Harvard-Smithsonian Center for Astrophysics, 60 Garden St, Cambridge, MA 02138 USA}
\altaffiltext{5}{Einstein Fellow}
\altaffiltext{6}{Department of Physics \& Astronomy, 4129 Frederick Reines Hall,
 University of California, Irvine, CA 92697-4575, USA}
\altaffiltext{7}{Department of Statistics, The University of Auckland, Private Bag 92019, 
Auckland 1142, New Zealand}
\altaffiltext{8}{Department of Physics and Astronomy, University of California, Los Angeles 90095}
\altaffiltext{9}{Center for Cosmology and AstroParticle Physics, The 
Ohio State University, Columbus, OH 43210, USA}
\altaffiltext{10}{Space Telescope Science Institute, 3700 San Martin Drive, Baltimore, MD 21218} 

\begin{abstract}
We present inferences on the geometry and kinematics of the broad-\Hbeta \ line-emitting region in four active galactic nuclei monitored as a part of the fall 2010 reverberation mapping campaign at MDM Observatory led by the Ohio State University. 
From modeling the continuum variability and response in emission-line profile changes as a function of time, we infer the geometry of the \Hbeta-emitting broad line regions to be thick disks that are close to face-on to the observer with kinematics that are well-described by either elliptical orbits or inflowing gas. We measure the black hole mass to be $\log_{10}(M_{\rm BH})~=~$~\mrkAmbh \ for Mrk\,335, \mrkBmbh \ for Mrk\,1501, \threecmbh \ for 3C\,120, and \pgmbh \ for PG\,2130+099. These black hole mass measurements are not based on a particular assumed value of the virial scale factor $f$, allowing us to compute individual $f$ factors for each target. Our results nearly double the number of targets that have been modeled in this manner, and investigate the properties of a more diverse sample by including previously modeled objects. We measure an average scale factor $\bar{f}$ in the entire sample to be log$_{10} \bar{f}$ ~=~\avgfwholesample \ when the line dispersion is used to characterize the line width, which is consistent with values derived using the normalization of the $M_{\rm BH}$--$\sigma$ relation. We find that the scale factor $f$ for individual targets is likely correlated with the black hole mass, inclination angle, and opening angle of the broad line region but we do not find any correlation with the luminosity. 
\end{abstract}

\section{INTRODUCTION}
\label{sec:introduction}
Over the past couple of decades, enormous improvements have been made
in our understanding of the physics behind active galactic nuclei
(AGNs) and their central engines. It is now widely accepted that AGNs
contain a supermassive black hole (BH) with some form of accretion
disk. The accretion disk photoionizes gas farther out in the
broad line region (BLR), from which we see emission lines that are
Doppler-broadened due to the motion of the gas around the BH. In
nearby quiescent galaxies, \mbh is measured using stellar and gas
dynamics (e.g., \citealt{McConnell13}), but the central regions of
AGNs and galaxies farther away are too small in angular extent to
allow such measurements. However, under the assumption that the motion
of the gas in the BLR of AGNs is dominated by the gravity of the BH,
we can directly measure the mass of the BH ($M_{\rm BH}$) by employing
reverberation mapping methods (e.g., \citealt{Blandford82};
\citealt{Peterson93}). Reverberation mapping (RM) makes use of the
variability of AGNs to determine the time delay between signals in the
continuum emission, which is thought to come from very close to the
BH itself, and the response of the gas in the BLR. As the delay is
due to the light-travel time between the central source and the BLR, a
measurement of the time delay between these signals yields a distance
of the BLR gas from the central source. Combined with a measurement of
the velocity field of the BLR gas, one can measure \mbh using the
so-called virial relation:
\begin{equation}
M_{\rm BH}~=~ \frac{fR\Delta V^2}{G}
\label{eq:eq1}
\end{equation}
where $R$ is the characteristic radius of the BLR, $\Delta V$ is the
line-of-sight velocity of the gas, and $f$ is a dimensionless scale factor
that accounts for the geometry, kinematics, and orientation of the BLR
itself.  All \mbh estimates outside the local universe are made using
the BLR in AGNs, making them powerful tools for exploring the BH
population across the observable universe.

In Equation~\ref{eq:eq1}, $R$ is measured via RM or via single-epoch
methods and $\Delta V$ is measured from the width of the emission
line. Because the BLR is currently
unresolvable, the true value of $f$ in each target is unknown, so an average scale factor $\bar{f}$ has commonly been used to
calculate \mbh in AGNs. Typically, it is assumed that AGNs follow the same
\msigma relation as quiescent galaxies and calculate the average scale factor $\bar{f}$ required to move the entire populaton of reverberation-mapped
AGNs onto the quiescent \msigma relation (e.g., \citealt{Onken04};
\citealt{Graham11}; \citealt{Park12}; \citealt{Grier13b}; \citealt{Woo15}; \citealt{Batiste17}). The unknown scale factor $f$ is the largest source of
uncertainty in AGN \mbh measurements today. Because AGNs are used
to characterize the BH population across the observable universe, it
is in our best interest to refine these measurements to be as
accurate as possible, and to do so involves the recovery of additional
information on the environment within the BLR to determine individual
scale factors in AGNs.

Until recently, RM efforts were typically only able to obtain
measurements of the average time delay between signals in the BLR and
the continuum --- because the BLR gas is not confined to one specific
radius, measuring the flux across an entire broad emission line yields
some characteristic radius of the line-emitting region. Cross
correlation methods are the most common way of obtaining this average
time delay (e.g., \citealt{Peterson04}), though other methods have been examined in the past (such as linear inversion; e.g., \citealt{Krolik95}). 
Recently, alternative
methods that model the light curves and/or transfer functions have begun
to be used (\citealt{Zu11}; \citealt{Grier12b}; \citealt{Skielboe15}; \citealt{Fausnaugh16a}). The
average time delay is sufficient to obtain a \mbh measurement, assuming $f$, but
does not yield information about the structure or kinematics within
the BLR itself. However, with sufficiently high-quality RM data, we
can actually resolve the time delays in velocity space and thus
recover information about the possible phase space structure within
the BLR. This velocity-resolved analysis has been successfully done for
several sets of data in recent years (\citealt{Bentz09a};
\citealt{Denney10}; \citealt{Barth11}; \citealt{Barth11b}; \citealt{Doroshenko12}; 
\citealt{Grier13a}; \citealt{Derosa15}; \citealt{Du16}; \citealt{Pei16}). In most cases, we see signatures of either
gas in bound elliptical orbits or inflowing gas, although signatures of
outflowing gas have also been seen (\citealt{Denney09c}; \citealt{Du16}).

To obtain more detailed information about the BLR, a few recent
studies have set out to recover the transfer function, or
velocity-delay map, that shows exactly how the variations in the
continuum emission are mapped into variations in the broad line
emission as a function of the line-of-sight velocity of the gas
(\citealt{Bentz10a}, \citealt{Grier13a}). These studies used maximum
entropy methods (MEM) implemented in a code called MEMECHO
(\citealt{Horne91}; \citealt{Horne94}), to recover the transfer
functions. In \cite{Grier13a}, hereafter referred to as Paper I, we
applied the MEMECHO code to five targets from a 2010 RM campaign
(described by \citealt{Grier12b}, hereafter G12) and successfully
recovered velocity-delay maps for four of the targets. The
velocity-delay maps confirmed the initial velocity-resolved time lag
results, also showing signatures of gas both in elliptical orbits and
inflowing, with possible disk-like or spherical geometries.  While the
MEMECHO velocity-delay maps help us to determine qualitatively what
kind of possible structures and kinematics we are seeing in the BLR,
they yield no concrete parameters on either the geometry or kinematics
of the BLR and ideally should be compared with models to make more
detailed and precise inferences.

Recently, other approaches have been developed to model reverberation
mapping datasets directly to obtain quantitative constraints on both
the geometry (\citealt{Li13}) and kinematics of the BLR (\citealt{Pancoast11};
\citealt{Brewer11}; \citealt{Pancoast14b}). Modeling both the geometry and kinematics yields a
measurement of \mbh independent of the scale factor $f$ that can be compared
with values returned by other RM methods. \cite{Pancoast14b} have
improved the flexibility of the BLR model used in their approach and
have successfully applied their methodology to model five AGNs from the LAMP
2008 RM campaign (\citealt{Bentz10a}; \citealt{Walsh09}) and
demonstrated its power to fit the data and provide constraints on the
BLR environment (\citealt{Pancoast14}; hereafter P14). 

In this study, we continue our investigation of the structure and
kinematics of the BLR that was begun in Paper I with the MEMECHO
velocity-delay maps. Here we aim to obtain constraints on the BLR
geometry and kinematics in our targets by directly modeling the data using the methods of \cite{Pancoast14b}. The aim
of this study is to further add to the sample of targets with
dynamical modeling information to learn about the RM population as a
whole. We also aim to compare the transfer functions recovered with
MEMECHO in Paper I to the information recovered from dynamical
modeling. In Section~\ref{sec:data}, we describe the data we used to model
the BLR and the spectral decomposition used to isolate the broad emission lines. 
In Section~\ref{sec:modelingmethods}, we briefly discuss the dynamical modeling method, and in Section~\ref{sec:results} we
describe the modeling results for each individual AGN. In Section~\ref{sec:discussion} 
we combine our results with those from the LAMP 2008 dataset (P14) to discuss the dynamical modeling sample as a whole, any trends found,
and calculate the mean scale factor $f$ for the sample. We conclude in Section~\ref{sec:summary} 
with a summary of our results and their implications.

\begin{deluxetable*}{lccccccccccr}
\tablewidth{0pt}
\tablecaption{Target Information}
\tablehead{
\colhead{Object} &
\colhead{RA} &
\colhead{DEC} &
\colhead{$z$} &
\colhead{A$_B$\tablenotemark{a} } & 
\colhead{N$_{\rm c}$\tablenotemark{b}} &
\colhead{N$_{\rm s}$\tablenotemark{c}} &
\colhead{N$_{\rm pix}$\tablenotemark{d}} &
\colhead{Wavelength}  & 
\colhead{$\tau_{\Hbeta}$\tablenotemark{e}} &
\colhead{$\sigma_{\rm line}$\tablenotemark{e}} &
\colhead{FWHM\tablenotemark{e}} \\
\colhead{ } &
\colhead{(J2000)} &
\colhead{(J2000)} &
\colhead{ } &
\colhead{(mag)} &
\colhead{ } &
\colhead{} &
\colhead{} &
\colhead{Range (\AA)} & 
\colhead{(days)} &
\colhead{(km s$^{-1}$)} &
\colhead{(km s$^{-1}$)} 
} 
\startdata
Mrk\,335 & 00 06 19.5 & +20 12 10 & 0.0258  &  0.153    	& 129  & 78 & 78 & 4800-4895   	& 14.1$^{+0.4}_{-0.4}$ &  1293 $\pm$ 64  & 1273 $\pm$ 64  	\\
Mrk\,1501 & 00 10 31.0 & +10 58 30 & 0.0893  &  0.422    	& 210  & 65 & 109 & 4800-4925   	& 15.5$^{+2.2}_{-1.8}$ &  3321 $\pm$ 107 & 3494 $\pm$ 35  	\\
3C\,120 & 04 33 11.1 & +05 21 16 & 0.0330  &  1.283    	& 192  & 69 & 83 & 4800-4900   	& 27.2$^{+1.1}_{-1.1}$ &  1514 $\pm$ 65  & 1430 $\pm$ 16    \\
Mrk 6 & 06 52 12.2 & +74 25 37 & 0.0188  &  0.585   		& 204  & 72 & 265 & 4725-5050  	& 9.2$^{+0.8}_{-0.8}$  & 3714 $\pm$ 68  & 2619 $\pm$ 24  	\\
PG\,2130+099 & 21 32 27.8 & +10 08 19 & 0.0630  &  0.192    	& 235  & 68 & 81 & 4800-4895		& 12.8$^{+1.2}_{-0.9}$ &  1825 $\pm$ 65  & 1781 $\pm$ 5   

 \enddata
\label{Table:tbl1}
\tablenotetext{a}{Galactic extinction values are from \cite{Schlegel98}}. 
\tablenotetext{b}{Number of epochs in continuum light curve.} 
\tablenotetext{c}{Number of spectral epochs.} 
\tablenotetext{d}{Number of pixels used in modeling analysis.}  
\tablenotetext{e}{These measurements were published by G12. $\tau_{\Hbeta}$ measurements were produced using {\tt JAVELIN}. $\sigma_{\rm line}$ was measured from the 
root-mean square residual spectrum, and FWHM was measured from the mean spectrum.}
\end{deluxetable*} 

\section{DATA PREPARATION}
\label{sec:data}

\subsection{Spectroscopic Data}
The spectra used in this analysis were taken during a RM campaign
carried out primarily at MDM Observatory in late 2010, hereafter referred to as AGN10. Details
on the data processing are discussed by G12. The spectra were obtained with the Boller and Chivens CCD spectrograph on the 1.3m
McGraw-Hill telescope over the course of 120 nights from 2010 August
31 to December 28. The continuum light curves consist of fluxes
measured from both spectroscopic and photometric observations; they
were constructed from data taken at multiple observatories, as
discussed by G12. General information on the five targets we examine
here is given in Table~\ref{Table:tbl1}. Prior to modeling, the spectra were calibrated to the
absolute flux of the narrow [O\,{\sc iii}]\,$\lambda5007$ emission
line using the procedure of \cite{vanGroningen92}. 

Table~\ref{Table:tbl1} presents some basic information on the data used in the modeling, including the number of spectral epochs and the number of epochs in each continuum light curve. In Table~\ref{Table:tbl1}, we
reproduce previous RM results for each target from G12 to compare with
the results from this study. These RM results from
G12 were all obtained using {\tt JAVELIN} to model the
continuum and light curves (see \citealt{Zu11}, \citealt{Zu13}, Paper I, or G12
for details).  

Paper I presents velocity-delay maps that include three emission lines: \Hbeta,  \Hgamma \ and \heii. However, the quality of the measurements for \heii \ and \Hgamma \ is lower than those for \Hbeta. The \heii \ emission line is broad, relatively weak, and very difficult to isolate from other components of the spectrum using spectral decomposition. The \Hgamma  \ line, while stronger, lies near the blue end of the wavelength coverage of our spectra, where there are no strong narrow emission lines. As such, the relative flux calibration for \Hgamma \ is of lower quality than that for \Hbeta. We thus restrict our analysis here to only the \Hbeta \ emission line, and defer discussion of additional emission lines to future RM studies with more favorable data quality for these line species. 

\subsection{Spectral Decompositon}
In both Paper I and G12, and indeed in most prior RM studies, the broad emission-line fluxes were initially measured by subtracting off a linear continuum underneath the emission line, measured using local continuum windows on either side of the emission line. However, several groups (e.g., \citealt{Barth11}; \citealt{Park12}; \citealt{Barth13}; \citealt{Hu15}) have developed methods to isolate various components of the AGN spectrum, allowing for the disentanglement of various broad emission line features from the rest of the AGN, such as the host galaxy starlight, \feii \ features and various other species that often blend with the emission lines we are investigating. The recent success with isolating these different components of the AGN spectrum allows one to measure light curves for various AGN components  despite strong starlight and blending (\citealt{Barth11}; \citealt{Barth13}). As a result, the spectral decomposition methods have been continually improved and implemented in RM spectra (\citealt{Park12}; \citealt{Barth15}). 

Most of our targets show complex features, significant \feii, and/or strong host galaxy starlight features in their spectra -- these issues affect the \Hbeta \ line profiles and thus have the potential to affect the resulting BLR models inferred. Thus, we opted to perform the spectral decomposition on our spectra to isolate the \Hbeta \ emission to allow us to subtract off all other spectral components than \Hbeta; this residual spectrum would then be used in our modeling. The spectral decomposition method used in our study is described in detail in \S 4.3 of \cite{Barth15}, though one modification was made: Instead of using 4th-order Gauss-Hermite functions to model the broad \Hbeta \ component and [\oiii], we use a 6th order Gauss-Hermite function because the line profiles in these objects were more complex than those examined by \cite{Barth15}. 
Because of the relatively small wavelength range of our MDM spectra, the fits were carried out over the entire range of the spectra, ranging from about 4200--5400~\AA \ in the rest frame (with small variations depending on the object redshifts). The model components include host starlight, a power law AGN continuum,  \ob,  \Hbeta, \HeII, \HeI, and \feii. The \Hbeta \ model includes both broad and narrow \Hbeta \ components as separate Gauss-Hermite models. We tried three different \feii \ templates from \cite{Boroson92}, \cite{Veroncetty04}, and \cite{Kovacevic10}, performing full decompositions with each template so we could compare the resulting \Hbeta \ profiles using each.  

We were unable to adequately fit the extremely complex line profile of Mrk 6, which has a very broad, asymmetric \Hbeta \ profile with significant stellar, \heii, and \feii \ emission all overlapping with \Hbeta \ as well. Because of the difficulty in isolating the \Hbeta \ emission in this target, we were unable to successfully produce a model from it, and it is thus excluded from further analysis. However, the three different \feii \ templates yielded remarkably consistent \Hbeta \ profiles in all four of our remaining sources, with the exception of some minor differences in the red wing of the \Hbeta \ emission. This region contains contributions from \feii \ and \hei, but the line profiles over the majority of the range spanning the \Hbeta \ line were nearly identical for each source. The one exception is Mrk\,1501; this target has a broader \Hbeta \ profile than the others, and the exact shape of the red wing depends more heavily on the \feii \ template used. We used the fits from the best \feii \ template of the three, chosen based on both visual inspection and the $\chi^2$ of the fit, for our modeling. In all four cases, the  \cite{Boroson92} template yielded the best fit to our data, though we again note that the differences were very minor and restricted to only the red wing of the \Hbeta \ profile. 

\begin{figure*}
\begin{center}
\includegraphics[scale = 0.5, angle = 0, trim = 0 0 0 0, clip]{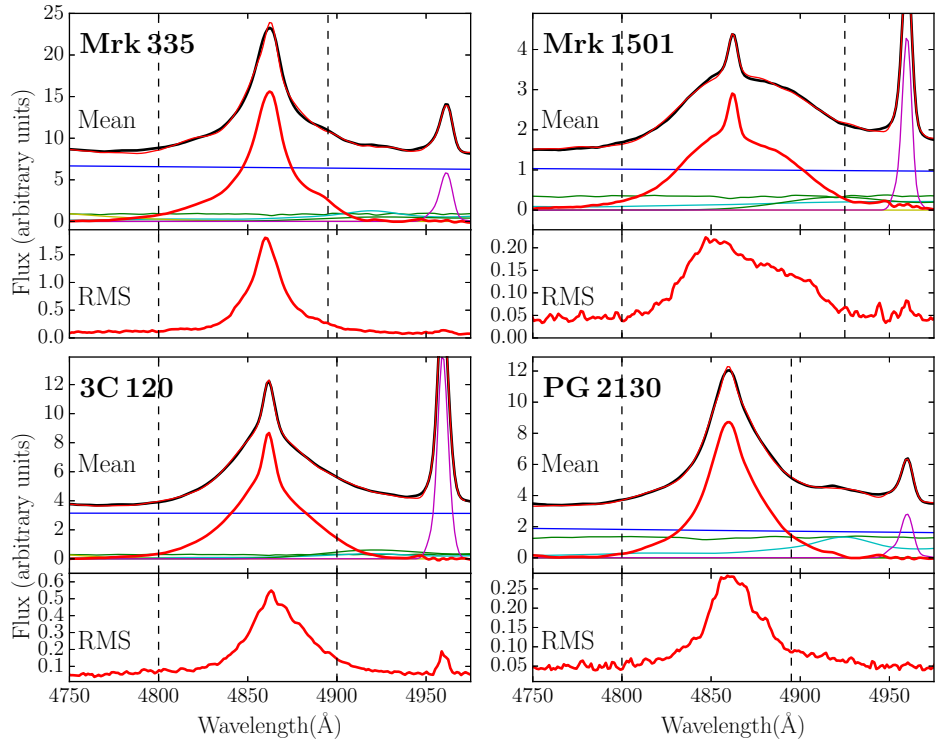}
\caption{Top subpanels: Spectral fits to the mean spectrum of each target. The data and full model are shown by the top black and red overlapping spectra.  The \Hbeta+\hei \ residual spectrum used in our dynamical modeling analysis is indicated by the thick red line for each target. Each component of the fit is shown individually as well: starlight (green), AGN power-law continuum (blue), \feii \ (cyan), \heii \ (yellow), \hei \ (green), and [\oiii] (magenta). Bottom subpanels: root-mean-square (RMS) residual spectra for each target, created from the spectral decompositions with all components subtracted off of \Hbeta \ except \hei \ as described in the text. The dashed vertical lines indicate the upper and lower limits to the wavelength region used in our modeling analysis.}
\label{fig:spec_decomp}
\end{center}
\end{figure*}

We show the various components of the spectral fits using the \cite{Boroson92} \feii \ templates in Figure~\ref{fig:spec_decomp} for the mean spectrum of each of our AGN. Because the \hei \ components often appeared degenerate with portions of the red wing of the \Hbeta \ emission and the \feii \ emission in that region, we opted to leave the possible \hei \ emission in our spectra to model the case where all of the flux in this region is due to \Hbeta. Our final spectra for the modeling analysis are thus the original data with all other components subtracted except for \hei. We note, however, that for our four objects, the contribution from \hei \ is small. 
To minimize possible systematics caused by differences between the \feii \ template fits, we include the red wing of the \Hbeta \ emission line only as far as the three templates were in general agreement on the \Hbeta \ line profile (see Figure~\ref{fig:spec_decomp} and Table~\ref{Table:tbl1} to see the exact wavelength ranges used in the modeling for each source). 

\section{DYNAMICAL MODELING METHOD}
\label{sec:modelingmethods}

We model individual RM datasets using a simply 
parameterized phenomenological modeling code for the BLR that is fully described
by \citet{Pancoast14b}.  
In addition, we have added the AGN redshift as a free
parameter with a narrow Gaussian prior of width 1\,\AA\ to account for imperfect
determination of the redshift from nearby narrow lines.
We also describe some systematic uncertainties in the model in Section \ref{modelcaveats}. 

The distribution of broad line emission in position and velocity space
is sampled using a number of massless point test particles that instantaneously and 
linearly reprocess the AGN continuum flux into broad line flux.  
The position of the point particles determines the time lag with which
the continuum flux is reprocessed and the velocity of the point
particles determines the Doppler shifted wavelength at which 
the line flux is emitted.  Given an input continuum light curve, we can use
the positions and velocities of the point particles to generate model
emission line profiles that can be directly compared with the data.

\subsection{Geometric Model}
\label{geomodel} 
We use a flexible model for the BLR geometry that parameterizes the
positions of the point particles using radial and angular
distributions.  For the radial distribution of point particles we use
a Gamma distribution
\begin{eqnarray}
 p(x|\alpha,\theta) \propto x^{\alpha-1} \exp \left( -\frac{x}{\theta} \right)
\end{eqnarray}
that generates profiles ranging from
Gaussian to exponential or steeper.  The Gamma distribution is offset from the origin by
the Schwarzschild radius, $R_s~=~2GM_{\rm BH}/c^2$ plus a minimum BLR radius, $r_{\rm min}$. The quantity $r_{\rm min}$ is measured relative to the radius at which the continuum emission is emitted; we here assume the continuum to be emitted at $r = 0$, but note that this assumption may not be correct (see Section~\ref{modelcaveats}). We assume that the outer edge of the BLR is small enough that length of the RM campaign is more than sufficient to measure all time delays within it;  we thus restrict the offset Gamma distribution to an outer radius $r_{\rm out}~=~c\Delta t_{\rm data}/2$, where $\Delta t_{\rm data}$ is the total time between 
the beginning of the continuum model light curve and the first epoch of the broad emission-line light curve. 
We perform a change of variables between
$(\alpha,\,\theta, r_{\rm min})$ and $(\mu,\,\beta,\,F)$ such that
\begin{eqnarray}
 \mu &=& r_{\rm min} + \alpha\theta \label{eqn_mu} \\
 \beta &=& \frac{1}{\sqrt{\alpha}} \label{eqn_beta}\\
 F &=& \frac{r_{\rm min}}{r_{\rm min}+\alpha\theta} \label{eqn_f}
 \end{eqnarray}
where $\mu$ is the mean radius, $\beta$ determines the shape of the
 Gamma distribution, and $F$ is a fractional radius  corresponding to
 $r_{\rm min}$/$\mu$. The radial distribution has a standard deviation 
 given by $\sigma_r~=~\mu \beta (1 - F)$.  As a part of the modeling process, we also calculate the
 mean radius $r_{\rm mean}$, median radius $r_{\rm median}$, mean time lag
 $\tau_{\rm mean}$, and median time lag $\tau_{\rm median}$ for specific realizations of point particle positions. 
 We allow the system to deviate from spherical by including an opening angle ($\theta_o$, defined as the
half-opening angle of the BLR disk) that allows the geometry to range from a razor-thin disk to a sphere, and also allow the system to be inclined towards the observer by the inclination angle ($\theta_i$).  Values of $\theta_o~ =~0\degree$ and $\theta_o~ =~90\degree$
correspond to a thin disk and spherical geometries, respectively; values of $\theta_i~=~0\degree$ and $\theta_i~=~90\degree$ correspond to
face-on and edge-on geometries, respectively. 

For additional flexibility, the BLR model also allows three different types of asymmetry: First, we allow for asymmetric line emission from each point particle. 
We weight the emission seen by the observer from each point particle as follows:
\begin{eqnarray}
 W(\phi)~=~\frac{1}{2} + \kappa \cos \phi. \label{eqn_kappa}
\end{eqnarray}
where $W$ is the weight given to each point
particle (between 0 and 1), $\phi$ is the angle between the observer's and point particle's line of sight to the
central source, and $\kappa$ is a parameter that allows for anisotropic emission from the
point particles. The quantity $\kappa$ ranges between $-0.5$ and 0.5: A value of $-0.5$  corresponds to the observer seeing more emission from the far side of
the BLR due to the point particles emitting preferentially back toward the continuum source, and a value of 0.5 corresponds to the observer seeing more line emission from the near side of the BLR, with the point particles preferentially emitting away from the central ionizing source. 
Second, we allow for the preferential emission from the outer faces of the disk by changing the angle for a point particle's
displacement from a flat to thick disk defined by 
\begin{eqnarray}
 \theta~=~{\rm cos^{-1}} [\cos \theta_o + (1 - \cos \theta_o)\times U^\gamma ]  
 \label{eqn_gamma}
\end{eqnarray}
where $U$ is a random number drawn uniformly between the values of 0
and 1. The $\gamma$ asymmetry parameter controls the extent to which BLR emission is
concentrated in the inner regions or the outer faces of the disk. Values of $\gamma$ range from 1 to 5, where $\gamma~=~1$ corresponds to uniform
concentrations of point particles in the disk  and $\gamma~=~5$ corresponds to more point particles
along the faces of the disk. 
The third asymmetry parameter is $\xi$, defined as 
two times the fraction of point particles below the disk mid-plane.
$\xi$ allows for the mid-plane of the BLR to range from transparent to opaque: For $\xi~=~0$,
the mid-plane is opaque, and as $\xi \to 1$, it becomes transparent.

\subsection{Dynamical Model} 
The kinematics of the BLR are parameterized in the plane of the radial
and tangential velocities of the point particles in the Keplerian potential of the BH (radiation pressure is presumed to be negligible.) We allow for a fraction of particles $f_{\rm
  ellip}$ with elliptical orbits drawn from a distribution centered around the
circular orbital velocity (near-circular elliptical orbits); 
$f_{\rm ellip}$~=~0 and $f_{\rm ellip}$~=~1 represents none of and all of the particles having near-circular elliptical orbits, respectively.
The remaining 1 - $f_{\rm ellip}$ fraction of particles are in inflowing ($0 < f_{\rm flow} < 0.5$) 
or outflowing ($0.5 < f_{\rm flow} < 1$) orbits drawn
from a distribution centered around the radial escape velocity. 
The angle $\theta_{\rm e}$ adds
flexibility to the dynamics by allowing the distributions for inflow
and outflow velocities in the plane of the radial
and tangential velocities of the point particles to be rotated towards
the circular orbit velocity (for a more thorough discussion of this, see Section
2.5 in \citealt{Pancoast14b}). As $\theta_{\rm e} \to 90$ degrees, the
inflow and outflow velocity distributions approach the distribution
for near-circular elliptical orbits,
so models with low $f_{\rm ellip}$ at $\theta_{\rm e}$ =
90 degrees are the same as models with high
values of $f_{\rm ellip}$.

We also allow for a small
addition to the point particle velocity vector from macroturbulence, given by 
\begin{eqnarray}
 v_{\rm turb}~=~\mathcal{N}(0,\sigma_{\rm turb}) |v_{\rm circ}|
\end{eqnarray}
where $\sigma_{\rm turb}$ is the 
standard deviation of the Gaussian distribution from which a randomly-oriented 
macroturbulent velocity component is drawn with a prior between 0.001 and 0.1 and $v_{\rm circ}$ is the circular orbit velocity. 

\subsection{Continuum Models and Implementation \label{sect_model_implementation}} 
\begin{figure*}
\begin{center}
\includegraphics[scale = 0.38, angle = 0, trim = 0 0 0 0, clip]{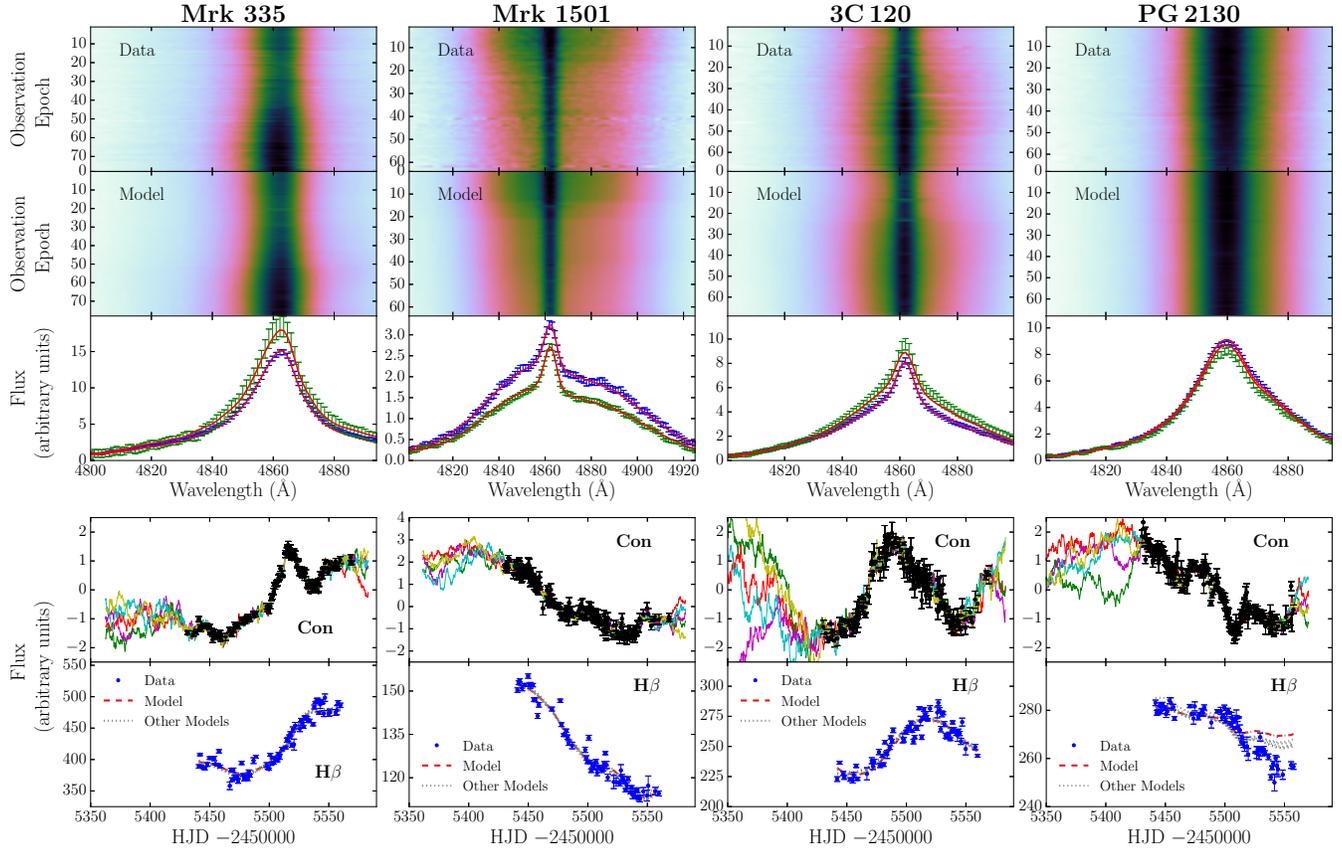}
\caption{Model fits to the continuum light curves and \Hbeta \ emission-line spectra for all four AGN. 
Top panels: The \Hbeta \ spectral time series of data used in the modeling. 
Second panels: One example of a model spectral time series, drawn randomly from the posterior PDF. 
Third panels: Two spectra drawn from the data (blue and green error bars) with a model fit to the spectra over-plotted in red. The model was drawn randomly from the posterior.
Fourth panels: Continuum light curves. The black points show the light curves from G12, and the various colors show random continuum models drawn from the posterior PDF from dynamical modeling.  
Bottom panels: The integrated \Hbeta \ emission-line light curve generated from the spectrally decomposed spectra used in the modeling analysis (blue). The light curves created from the simulated spectra in the middle panels are shown in red, and more models randomly drawn from the posterior PDF are shown in gray. 
}
\label{fig:all-rainbow}
\end{center}
\end{figure*}

In addition to a model for the BLR geometry and kinematics, we must also
model the AGN continuum light curve in order to evaluate the continuum
flux at arbitrary times for calculation of the broad line flux.  We
use Gaussian processes to model the stochastic AGN continuum
variability and interpolate between the continuum flux data points,
since it has been found to be a good model for larger samples of AGN
(e.g., \citealt{Kelly09}, \citealt{MacLeod10};
\citealt{Zu11}; \citealt{Fausnaugh16a}; \citealt{Kozlowski16}).  Using this model we can incorporate the uncertainty
in the interpolation into our constraints on the BLR geometry and
kinematics as well as extrapolate beyond the ends of the data to
evaluate the line flux from point particles with long time lags. We show examples of continuum 
Gaussian process models for each of our targets in the middle panels of Figure~\ref{fig:all-rainbow}. 

We pose this problem of fitting a model of the BLR to a RM dataset in terms of Bayesian inference and use Diffusive
Nested Sampling (\citealt{Brewer10}) 
of the BLR and AGN continuum model parameters.  Diffusive nested
sampling also allows for model comparison by calculating the
``evidence" value that normalizes the posterior PDF.  We compare the
time series of \Hbeta\ emission line profiles from the data with the
time series of model line profiles using a Gaussian likelihood
function.  In general, the model cannot match the data completely to
within the small quoted uncertainties and the likelihood function must
be softened by dividing the logarithm of the likelihood by a
temperature $T$, where $T \ge 1$.    
This is equivalent to multiplying the spectral 
uncertainties in the Gaussian likelihood function by $\sqrt(T)$. Using larger values of the
temperature incorporates additional uncertainty into the likelihood
function, which can be thought of as due to underestimated spectral
flux errors or the use of a BLR model that does not include sufficient
flexibility to match all features in the data.

Given the high-dimensional parameter space and 
high quality of the data, it is important to check the 
convergence of the BLR model inference.  We can improve convergence by reducing the 
numerical noise of the model emission line profiles, using 2000 point particles and drawing 10
velocities for each.  This results in numerical noise from changes in the model line profile for fixed BLR
model parameters that is on the order of the spectral uncertainties in the data, $\sigma_{\rm spectra}$.  However,
given that we use temperature values greater than one 
($T=12-20$ for Mrk 335, $T=5$ for Mrk 1501, $T=5-7$ for 3C 120, and $T=35-55$ for PG2130+099),
the numerical noise is smaller than the effective spectral errors, 
$\sigma_{\rm effective} = \sigma_{\rm spectra}\sqrt(T)$, for all but one epoch out of 275 for the four AGN.
As described in Section~\ref{sec:pg2130}, the higher temperature for PG2130+099 is due to a poor model fit at the end of the light curve. 
We test for convergence by running each AGN multiple times using different starting parameter values
and comparing the results.  
The final posterior PDFs for each AGN are then created by 
adding together an equal number of samples from each of the runs.

\section{RESULTS FOR INDIVIDUAL OBJECTS}
\label{sec:results} 
We here present the detailed dynamical modeling results for our sample
of four AGNs.  For each source, we discuss the quality of the model fit to the data,
the constraints on the geometry and kinematics of the BLR, and the shape
of the transfer function.  When possible, we compare
these constraints from dynamical modeling to the results from the
analysis in G12 and the MEMECHO analysis presented by \cite{Grier13a}.  The
posterior median and 68\% credible intervals for the BLR model
parameters of each target are summarized in Table~\ref{Table:tbl2},
while the individual $f$ values inferred for each target are listed in Table~\ref{Table:tbl3}. We also 
provide histograms of the posterior distributions for a few of the 
most significant model parameters for each object in an Appendix. 

\begin{deluxetable*}{lcccccc} 
\tablewidth{0pt} 
\tablecaption{Dynamical Model Parameters} 
\tablehead{ 
\colhead{Parameter} & 
\colhead{Mrk 335} & 
\colhead{Mrk 1501} & 
\colhead{3C\,120} & 
\colhead{PG\,2130+099 } 
}  
\startdata 
$r_{\rm out}$ (light-days)\tablenotemark{*} & 39.1796 & 39.6777 & 54.2524 & 50.1817  \\ 
$r_{\rm mean}$ (light days) &   $17.98^{+ 1.74}_{- 2.16}$ &   $17.86^{+ 1.16}_{- 1.20}$ &   $23.31^{+ 0.99}_{- 0.96}$ &   $13.51^{+ 3.26}_{- 2.79}$ \\ 
$r_{\rm median}$ &   $16.91^{+ 2.10}_{- 2.68}$ &   $16.44^{+ 1.74}_{- 1.61}$ &   $21.59^{+ 1.30}_{- 1.39}$ &   $ 9.13^{+ 4.34}_{- 3.05}$ \\ 
$r_{\rm min}$ (light days) &   $ 1.28^{+ 2.33}_{- 0.97}$ &   $ 5.33^{+ 4.34}_{- 3.59}$ &   $ 0.90^{+ 1.51}_{- 0.70}$ &   $ 1.32^{+ 0.81}_{- 0.76}$ \\ 
$\sigma_{r}$ (light days) &   $22.33^{+ 7.89}_{- 6.42}$ &   $16.17^{+ 9.02}_{- 5.65}$ &   $42.84^{+ 5.16}_{- 7.17}$ &   $30.12^{+16.79}_{-11.59}$ \\ 
$\tau_{\rm mean}$ (days)  &   $18.86^{+ 1.81}_{- 2.34}$ &   $17.08^{+ 1.03}_{- 1.21}$ &   $23.84^{+ 1.01}_{- 0.92}$ &   $13.22^{+ 3.44}_{- 2.87}$ \\ 
$\tau_{\rm median}$ &   $16.38^{+ 2.19}_{- 2.88}$ &   $14.96^{+ 1.22}_{- 1.41}$ &   $20.62^{+ 1.08}_{- 1.08}$ &   $ 7.79^{+ 3.97}_{- 2.57}$ \\ 
$\beta$ &   $ 0.85^{+ 0.15}_{- 0.14}$ &   $ 0.87^{+ 0.38}_{- 0.23}$ &   $ 0.94^{+ 0.07}_{- 0.06}$ &   $ 1.34^{+ 0.21}_{- 0.22}$ \\ 
$\theta_o$ (degrees) &   $38.1^{+ 4.7}_{- 5.2}$ &   $21.7^{+12.1}_{- 6.4}$ &   $21.1^{+ 8.0}_{- 5.2}$ &   $33.0^{+12.1}_{-12.2}$ \\ 
$\theta_i$ (degrees) &   $35.3^{+ 4.5}_{- 4.8}$ &   $20.5^{+ 5.0}_{- 5.7}$ &   $17.6^{+ 5.4}_{- 3.3}$ &   $30.2^{+11.0}_{-10.1}$ \\ 
$\kappa$ &   $-0.49^{+ 0.01}_{- 0.00}$ &   $-0.19^{+ 0.14}_{- 0.15}$ &   $-0.43^{+ 0.11}_{- 0.06}$ &   $-0.33^{+ 0.29}_{- 0.11}$ \\ 
$\gamma$ &   $ 4.66^{+ 0.25}_{- 0.53}$ &   $ 1.99^{+ 1.68}_{- 0.86}$ &   $ 2.25^{+ 1.61}_{- 0.89}$ &   $ 3.61^{+ 1.00}_{- 1.41}$ \\ 
$\xi$ &   $ 0.32^{+ 0.11}_{- 0.11}$ &   $ 0.57^{+ 0.13}_{- 0.20}$ &   $ 0.72^{+ 0.19}_{- 0.19}$ &   $ 0.44^{+ 0.31}_{- 0.20}$ \\ 

$\log_{10}(M_{\rm BH}/M_\odot)$ &   $ 7.25^{+ 0.10}_{- 0.10}$ &   $ 7.86^{+ 0.20}_{- 0.17}$ &   $ 7.84^{+ 0.14}_{- 0.19}$ &   $ 6.92^{+ 0.24}_{- 0.23}$ \\ 
$f_{\rm ellip}$ &   $ 0.02^{+ 0.03}_{- 0.01}$ &   $ 0.40^{+ 0.23}_{- 0.19}$ &   $ 0.56^{+ 0.18}_{- 0.20}$ &   $ 0.15^{+ 0.34}_{- 0.12}$ \\ 
$f_{\rm flow}$ &   $ 0.25^{+ 0.16}_{- 0.17}$ &   $ 0.26^{+ 0.17}_{- 0.18}$ &   $ 0.25^{+ 0.18}_{- 0.17}$ &   $ 0.27^{+ 0.18}_{- 0.18}$ \\ 
$\theta_e$ (degrees) &   $23.77^{+10.33}_{-13.40}$ &   $22.77^{+21.31}_{-14.79}$ &   $12.30^{+16.27}_{- 8.67}$ &   $18.06^{+21.63}_{-13.03}$ \\ 
$\sigma_{\rm turb}$ &   $ 0.007^{+ 0.020}_{- 0.005}$ &   $ 0.076^{+ 0.016}_{- 0.024}$ &   $ 0.070^{+ 0.019}_{- 0.026}$ &   $ 0.052^{+ 0.035}_{- 0.047}$ 
\enddata 
\tablenotetext{*}{$r_{\rm out}$ is the maximum allowed distance of the point particles from the origin (see Section \ref{geomodel}); it is the only parameter listed here that is set ahead of time and not calculated by the model. } 
\label{Table:tbl2} 
\end{deluxetable*}  

\begin{deluxetable}{lcc} 
\tablewidth{0pt} 
\tablecaption{Inferred $f$ from Dynamical Modeling} 
\tablehead{ 
\colhead{Object} & 
\colhead{log$_{10}(f_{\rm \sigma})$} & 
\colhead{log$_{10}(f_{\rm FHWM})$}  
}  
\startdata 
Mrk 335 &   $0.59^{+0.10}_{-0.10}$ &   $0.60^{+0.10}_{-0.10}$ \\ 
Mrk 1501 &   $0.34^{+0.20}_{-0.17}$ &   $0.30^{+0.20}_{-0.17}$ \\ 
3C120 &   $0.76^{+0.14}_{-0.19}$ &   $0.81^{+0.14}_{-0.19}$ \\ 
PG2130+099 &   $0.00^{+0.24}_{-0.23}$ &   $0.02^{+0.24}_{-0.23}$ 
\enddata 
\tablecomments{Virial products used to calculate $f$ were determined using $\sigma$ measured from the RMS residual spectrum and FWHM from the mean spectrum, as presented in Table~\ref{Table:tbl1}.} 
\label{Table:tbl3} 
\end{deluxetable}  


\subsection{Mrk\,335}
\label{sec:mrk335sec}
Mrk\,335 is a narrow-line Seyfert 1 galaxy that has been observed in
several RM campaigns (\citealt{Kassebaum97}; \citealt{Peterson98};
\citealt{Peterson04}; G12).  We use a number of
different comparisons between the model and the data to illustrate the
quality of the model fit in this target.  First, we show the changing
shape of the \Hbeta\ emission-line profile as a function of time for
both the model and the data, as seen in the top two panels of Figure~\ref{fig:all-rainbow} for a model drawn randomly from the
posterior PDF. We also show two examples of the \Hbeta\ emission-line profile in the middle panel of the left column in
Figure~\ref{fig:all-rainbow} to see how well the model fits the
detailed line shape. Finally, we compare the integrated emission-line flux for
the same model along with the data (shown in the lowest panel in 
Figure~\ref{fig:all-rainbow}) to illustrate how well the model
matches the overall variability in the data.  The BLR model fits the overall variability of
the \Hbeta\ emission and the detailed line shape in Mrk\,335 very well. 

\begin{figure}
\begin{center}
\includegraphics[scale = 0.4, angle = 0, trim = 0 0 0 0, clip]{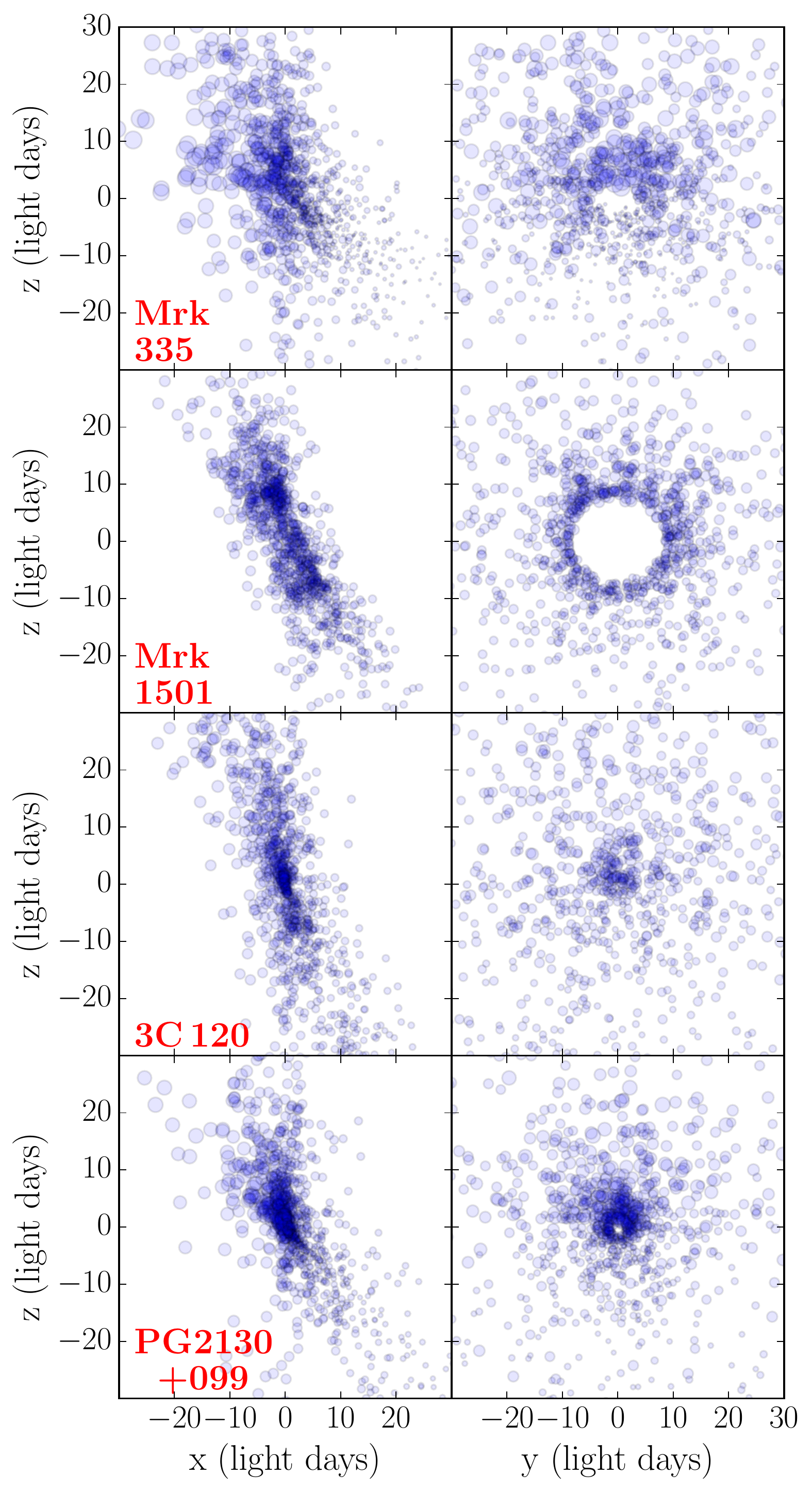}
\caption{Geometry of the \Hbeta-emitting region for all four sources. We show
  a model of the BLR drawn randomly from the posterior distribution. The left
  subpanels show an edge-on view of the BLR, along the $y$ axis, while
  the right subpanels show the BLR from the observer's point of view from
  along the positive $x$ axis. Each point corresponds to a point
  particle in our BLR model. The size of the points is proportional to the
  relative amount of \Hbeta \ emission coming from each particle when
  exposed to the same continuum flux.  
   }
\label{fig:all-geo}
\end{center}
\end{figure}

To illustrate the possible geometries of the BLR, we show the geometry of a model randomly drawn from the posterior in Figure
\ref{fig:all-geo}. By examining the inferred parameters in Table~\ref{Table:tbl2} and the posterior distributions of each parameter, we find that Mrk\,335 is best described by a thick disk 
with preference for more emission at the faces of the disk, more emission at the far side of the BLR, and a multi-modal distribution with solutions allowing for a mostly transparent or mostly opaque disk mid-plane. The radial Gamma distribution shape
parameter corresponds to a distribution with a tail that is between exponential and gaussian. 
The measured mean and median time lags are slightly higher than the time lag measured by \cite{Grier12a} using {\tt JAVELIN}; see Table~\ref{Table:tbl1}.  
However, we note that the model parameters $\tau_{\rm mean}$ and $\tau_{\rm median}$ are difficult to compare directly to previously-published time delays, as {\tt JAVELIN} assumes a simple top hat transfer function when modeling the light curves, which is very different than the transfer functions recovered here and in Paper I.

There are two types of dynamical solutions dominating the posterior PDF for Mrk\,335, as can be seen by the multiple peaks visible in the posterior distribution of the kinematic parameters shown in the Appendix. The first solution puts most of the point particles in near-circular elliptical orbits with the remaining particles being in radial inflowing orbits. The second 
solution has almost no near-circular elliptical orbits, with the inflowing orbits having a larger component of tangential velocity compared to radial, making them more similar to elliptical orbits. While the kinematics model also allows for the possibility of macroturbulent
velocities, we find that macroturbulent velocities do not contribute
significantly to the dynamics in Mrk\,335.

\begin{figure*}
\begin{center}
\includegraphics[scale = 0.38, angle = 0, trim = 0 0 0 0, clip]{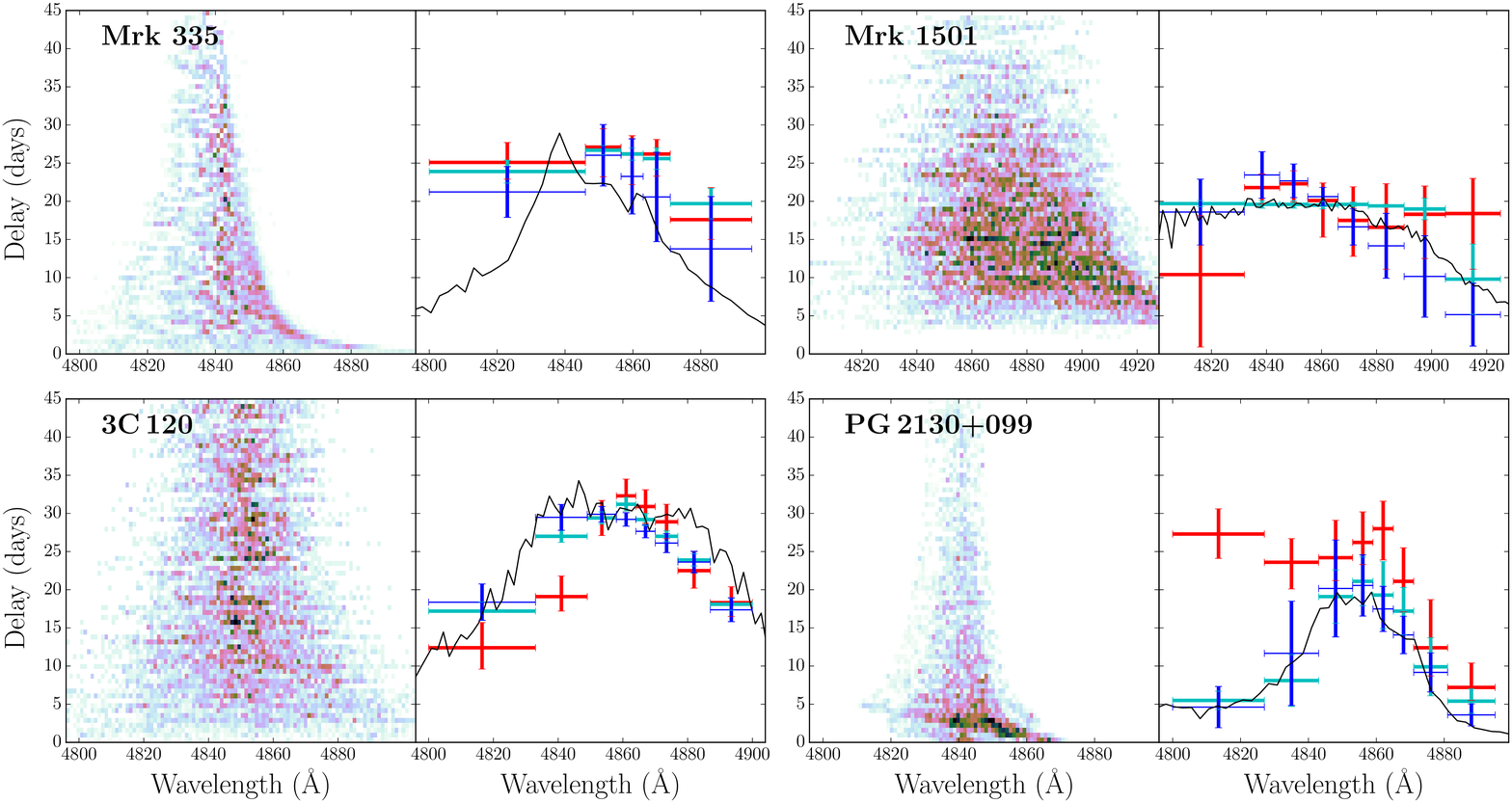}
\caption{Transfer functions and velocity-resolved time delays for all four AGN. For each source, the left panel shows 
  a representative transfer function drawn from the posterior PDF. The right panel for each source shows the velocity-resolved time delays for a number of wavelength bins. The solid black line shows the mean time delays computed by the model corresponding to the transfer function shown, and the blue crosses show the median lag values recovered from all of the model fits. To compare the model lags (blue and black) with the lags measured via cross correlation techniques, we also show mean time delays measured via cross correlation of the continuum light curves with light curves generated from the decomposed spectra (red crosses). We also cross correlated the model spectra from all of the samples with the continuum light curve and measured the median time lag in each bin (cyan crosses). }
\label{fig:all-veldel}
\end{center}
\end{figure*}

A velocity-delay map was recovered for the \Hbeta \ emission line in
Mrk\,335 and presented in Paper I. 
This velocity-delay map is not well-resolved, but shows a hint of asymmetric structure
that is consistent with inflowing gas (higher lags towards the blue and
shorter lags towards the red). The majority of the signal in the
velocity-delay map arises at low velocities at a range of radii from
about 15-40 light-days. We show a sample velocity-delay map from our model 
drawn from the posterior PDF in Figure
\ref{fig:all-veldel}; however, this transfer function is difficult to
compare to a MEMECHO velocity-delay map due to differences in
resolution. We do see much of the signal at similar radii (between
10-40 light-days) in both sets of maps. We also
compare the velocity-resolved RM lag measurements from Paper 1 to the
mean values of the inferred transfer functions from dynamical modeling
for the same wavelength bins, as shown in the right panel showing Mrk\,335 in
Figure~\ref{fig:all-veldel}.  The velocity-resolved lag
measurements based on the models agree to within the uncertainties with those measured from the decomposed spectra (red). We note that the original velocity bins used in Paper I are different than those used in this work; we updated the bins due to the changes in wavelength ranges used in the modeling, and thus the velocity-binned results differ somewhat from Paper I. We show the original bins and measurements from Paper I in gray for comparison, but note that there are some visible differences due to the fact that we are using decomposed spectra. 

The black hole mass in Mrk\,335 
is in agreement with the
previous measurement of log$_{10}$($M_{\rm BH}$)~=~7.29~$\pm$~0.05,
from the G12 analysis (all \mbh measurements from the G12
data have been updated using log~$f$~=~0.63, corresponding to $f$~=~4.31, following
\citealt{Grier13b}). Note that the uncertainties in \mbh from G12 only
take into account the measurement errors in $\tau$ and $\sigma$ used
to calculate the virial product --- the uncertainties do not include
the uncertainty in the scale factor $f$, which is estimated to be $\sim 0.4$
dex. 

\subsection{Mrk\,1501}
\label{sec:mrk1501}  

Our data for Mrk\,1501 constitute the first RM data set for this
target, and our model of the BLR was once again able to fit the spectral
variations quite well, even fitting the broad, asymmetric shape of
\Hbeta\ (Figure~\ref{fig:all-rainbow}). Like Mrk~335, the geometry of Mrk\,1501 is found to be a thick
inclined disk. Unlike Mrk\,335, however, the distribution of point particles in the BLR of Mrk\,1501 is inferred to be fairly close to uniform throughout the disk,
although solutions with more point particles at the edges of the disk are not ruled out. We
see a preference for emission from the far side of the BLR
rather than the near side and mid-plane of the disk to be partially transparent.
We show an example of a possible geometry in Figure~\ref{fig:all-geo}.
The radial profile of \Hbeta \ emission has a
Gamma distribution shape parameter that is inferred to be close to
exponential and the inferred mean and median time delays are 
consistent with values found by G12.

For Mrk\,1501, we find the kinematics to be a combination of elliptical and inflowing orbits where the fraction of elliptical orbits is not well-constrained but the remaining orbits are strongly preferred to be inflowing with a range of radial to tangential orbits. While the contribution of macro turbulent velocities is not great, it approaches the maximum value of 0.1 allowed by the prior and thus in this case could be larger but may be limited by the prior. 

Velocity-binned results from Paper I indicate inflowing gas, with
higher time lags towards the blue and lower time lags towards the
red. A somewhat blurred velocity-delay map recovered using MEMECHO
shows the same signature. As expected
from the inferred parameters, we see that the transfer functions
recovered from the dynamical modeling
(Figure~\ref{fig:all-veldel}) also show strong signatures of
inflowing gas. The velocity-binned results shown in the right Mrk\,1501
panel of Figure~\ref{fig:all-veldel} show consistent features,
although the mean time delays measured in the central velocity bins
from the model (shown in blue) and in light curves created from the decomposed spectra (red) are somewhat higher on average than the
time delays found using the RM techniques in Paper I (shown in gray).

We infer a black hole mass that is slightly lower than the previous measurement from the G12 data of log$_{10}$($M_{\rm BH}$)~=~
8.16~$\pm$~0.06, suggesting that the scale factor $f$ for this source deviates from the value (log $f$ = 0.63) that is commonly assumed.
We note that there is a degeneracy between the black hole mass, inclination angle, and
opening angle of the system, which limits the precision to which \mbh
can be measured with this approach (see Figure~\ref{fig:2dposteriors} in the Appendix).  This degeneracy arises because the model is trying to
match the width of the \Hbeta\ emission-line profile, and all three of
these parameters affect the measured line width. For a thin disk,
viewing the BLR closer to face-on will decrease the measured line
width, while increasing the opening angle of the disk will increase
the measured line width. While this degeneracy is partially broken by the transfer function, an independent method for measuring the
inclination angle or opening angle of the BLR would allow for an even more
precise measurement of the black hole mass (see Section~\ref{sec:threec}). 

\subsection{3C\,120}
\label{sec:threec}
3C\,120 is a well-studied radio-loud galaxy that has been observed in
multiple RM campaigns (e.g., \citealt{Peterson98},
\citealt{Peterson04}). Our model of the BLR was able to fit the
spectral variations and line shape from the 2010 data set 
quite well (Figure~\ref{fig:all-rainbow}).  We again see a preferred thick disk geometry in 3C\,120
and the inclination angle of the system is well-constrained at $\theta_{\rm
  i}$~=~\threecinclination \ relative to the observer. 3C\,120 is of
particular interest because there are external indicators of
inclination angle in the system obtained from radio jet orientation, which has shown to be linked to the BLR rotation axis 
(e.g., \citealt{Wills86}; \citealt{Marscher02}; \citealt{Jorstad05};
\citealt{Agudo12}). \cite{Marscher02} first determined the upper limit
on the jet viewing angle to be 20 degrees, and further work by
\cite{Jorstad05} measured a viewing angle of 20.5~$\pm$~1.8
degrees. Later work by \cite{Agudo12} determined a jet viewing angle
of $\theta$~=~16 degrees and we estimate the uncertainty in this
measurement from their paper to be about $\pm$ 3 degrees. Our $\theta_{\rm i}$
measurement from dynamical modeling is both well-constrained and
consistent with these measurements, indicating that in this system, the BLR orientation and jet orientation are aligned. 

We see a preference for emission from the far side of the disk
and for a mostly transparent disk mid plane, 
 although whether the emission is distributed equally throughout the disk or concentrated at the faces of the disk is not well-constrained. 
 Figure~\ref{fig:all-geo} shows an example of a possible geometry of the \Hbeta-emitting BLR in 3C\,120.
The radial distribution of \Hbeta\ emission in 3C\,120 has a Gamma distribution shape
parameter that is close to exponential.
We obtain a mean and median time delay 
consistent to within the uncertainties with the values reported by G12 
as well as recent work by \cite{Kollatschny14}, who report $\tau_{\Hbeta}$ = 27.9$^{+7.1}_{-5.9}$ days.

The kinematics of 3C\,120 are inferred to be a combination of near-circular elliptical
orbits and inflow on mostly radial orbits.
These kinematics are consistent with those recovered from velocity-delay maps using MEMECHO, which are the cleanest of the entire sample and show signatures consistent with those
expected from elliptical orbits in an inclined disk or a spherical
shell, with the \HeII \ emission line showing signs of inflowing gas. 
\cite{Kollatschny14} also performed a velocity-resolved analysis of 3C\,120 RM data from a
separate campaign and found similar features. 
The velocity-binned mean time delays
shown in Figure~\ref{fig:all-veldel} are also consistent with those measured in Paper I, though the measurements made directly from the spectra (red) deviate somewhat from the model (blue) in the second-bluest wavelength bin.

We infer a value for the black hole mass 
to be consistent with measurements from G12, who report
log$_{10}$($M_{\rm BH}$)~=~ 7.72 $\pm$ 0.04. 
As found in Mrk\,1501, we see a strong correlation between $M_{\rm BH}$, inclination angle, and opening angle
for this object. The inclination measurement for 3C\,120 made using radio jet orientation can also provide additional external constraints on \mbh that, if \mbh were less well-constrained by the model, could be used to narrow down the black hole mass further. However, the uncertainties in the jet inclination angle measurement are not well-constrained and are likely similar to the inclination angle uncertainties inferred by the model, so in this case, considering the jet inclination angle would not result in a substantial increase in precision in $M_{\rm BH}$. We note, however, that in other objects for which \mbh is less well-constrained, the additional information provided by external measurements such as radio jet inclinations could significantly improve precision in \mbh measurements, assuming the radio jets and BLR axes are aligned in all cases. 

\subsection{PG2130+099} \label{sec:pg2130}
PG\,2130+099 is a narrow-line Seyfert 1 galaxy that has been the target of several RM campaigns (\citealt{Kaspi00}, \citealt{Grier08},
\citealt{Grier12b}). Figure~\ref{fig:all-rainbow} shows our model fit to the PG\,2130+099 spectral time series. Overall, the model
was able to reproduce the detailed spectral shape well, but was unable to reproduce the integrated line variability during the last third of the campaign, possibly due to the low levels of variability throughout the campaign. It is also possible that the mismatch occurs because of a non-linear response by the \Hbeta \ emission line (such behavior has been reported, for example, in NGC\,5548 by \citealt{Goad16}) and thus this particular model will not provide an optimal fit. 
Similar to our other targets, PG\,2130+099 is well-described as a somewhat-inclined 
thick disk. We find preferential emission from the far side of the disk 
and more emission from the faces of the disk, 
although the transparency of the disk mid-plane is not well-constrained. 
Figure~\ref{fig:all-geo} shows an example of a possible geometry for the \Hbeta-emitting region of this target. The radial distribution of \Hbeta \ emission has a well-constrained Gamma distribution shape parameter corresponding to profiles steeper than exponential. We obtain mean and median time delays consistent with the value measured by G12.

The kinematics of the BLR are dominated by inflowing orbits, with a combination of radial and tangential velocities,  
although solutions with all near-circular elliptical orbits are not entirely ruled out. We do find a small contribution to the dynamics from
macroturbulence. The velocity-delay maps recovered by these models (Figure~\ref{fig:all-veldel}) show asymmetry reminiscent of those shown in Paper I; however, it appears that much of the emission is concentrated at smaller time delays, and there is stronger symmetric structure in the model velocity-delay maps that is more indicative of near-circular elliptical orbits. We also see that the two bluest velocity bins show much faster responses in the model than they do in the light curves created from the decomposed spectra (and also in Paper I). 

The black hole mass in PG\,2130+099 is is significantly lower than 
the mass calculated from G12 of log$_{10}$($M_{\rm BH}$)~=~ 7.56~$\pm$~0.04. This suggests that the true scale factor $f$ in this system is different than the average value for the reverberation-mapped sample, again indicating the importance of individual scale factors when considering individual AGNs. 

\section{DISCUSSION}
\label{sec:discussion}

\subsection{Overview of Geometric and Kinematic Results}
Overall, many of the BLR geometric and kinematic model parameters are
well-constrained for the four sources in our sample.  We find the BLR
geometries are best described by thick disks that are inclined such
that they are close to face-on for the observer ($\theta_{\rm i} < 45$; this is unsurprising given that these are Type 1 AGN). 
In all four sources, we find results that are consistent with preferential emission from the
far side of the disk, though this is not always well-constrained by
the model. The modeling results are generally consistent with previous
analyses: We find that the BLR radii and time delays are mostly 
consistent with those measured by G12, with some minor deviations. 

\begin{figure}
\begin{center}
\includegraphics[scale = 0.45, angle = 0, trim = 0 0 0 0, clip]{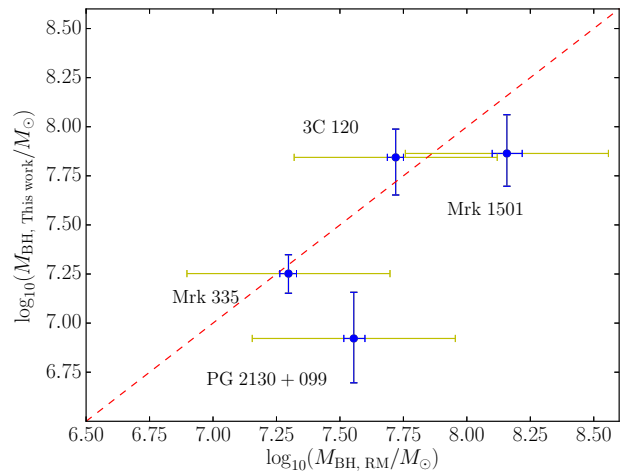}
\caption{\mbh computed by dynamical modeling (this work) versus \mbh
  computed using typical RM methods using time delays and line widths
  from G12 and log $\bar{f}$~=~0.63, as measured by
  \cite{Grier13b}. The red dashed line shows the identity. Blue
  points and error bars show the measurements with their published
  uncertainties; the yellow error bars in the background show
  uncertainties of 0.4 dex in the RM measurements, which is closer to
  that expected when taking into account the uncertainty in the scale factor $f$.}
\label{fig:mbh-comparison}
\end{center}
\end{figure}

We also compare the \mbh measurements from dynamical
modeling to those recovered from the RM analysis from G12, shown in
Figure~\ref{fig:mbh-comparison}. The previous measurements of \mbh
(hereafter referred to as $M_{\rm BH,RM}$) were again calculated assuming a
value for $\log f$ of 0.63 ($f$~=~4.31), following \cite{Grier13b}. As discussed above,
the uncertainties quoted for these $M_{\rm BH,RM}$ measurements
include only the measurement uncertainties in $\tau$ and $\sigma$ ---
they do not include the uncertainty in $f$ that is introduced by using
$\bar{f}$ calculated from the \msigma relation, which is
estimated to be $\sim$0.4 dex (shown as yellow error bars in Figure
\ref{fig:mbh-comparison}). With the exception of PG\,2130+099, the \mbh measurements from G12 are
consistent with those from dynamical modeling when this additional
uncertainty in $f$ is also taken into account. The mass inferred for PG\,2130+099 is marginally inconsistent with that reported by G12, suggesting that the scale factor $f$ in this object is lower than the average scale factor generally adopted for 
the entire AGN population, or that the model does not contain enough flexibility to fit the data sufficiently in this case. Given the relatively poor fit of the
PG\,2130+099 model light curves to the data at the end of the campaign (see Figure \ref{fig:all-rainbow}), we suspect that the latter is a likely scenario.  

\begin{figure}
\begin{center}
\includegraphics[scale = 0.46, angle = 0, trim = 0 0 0 0, clip]{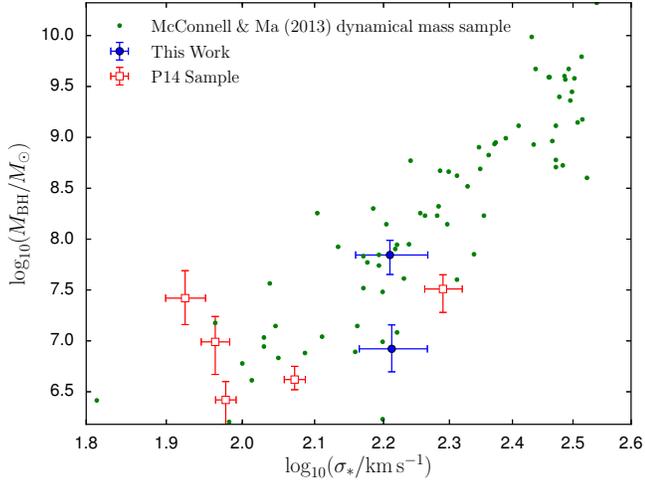}
\caption{log($\sigma_{*}$) versus log($M_{\rm BH}$). The small green squares are measurements of quiescent galaxies by \cite{McConnell13}. Open red squares show the five targets from P14 with \mbh measured via dynamical modeling, and closed blue circles represent the two objects in our study with dynamical modeling \mbh from this work and prior $\sigma_*$ measurements. The higher blue point shows 3C\,120, and the lower point shows PG\,2130+099. }
\label{fig:m-sigma}
\end{center}
\end{figure}

P14 show their \mbh measurements from dynamical modeling on the \msigma \ relation of quiescent galaxies with the dynamical \mbh measurements presented by \cite{McConnell13}, showing that the \mbh measured via dynamical modeling places their five sources in positions consistent with the quiescent \msigma \ relationship. Unfortunately, stellar velocity dispersion ($\sigma_*$) measurements exist only for two of our targets: 3C\,120, with $\sigma_*$~=~162~$\pm$~20~\kms\ (\citealt{Nelson95}), and PG\,2130+099, with $\sigma_*$~=~163~$\pm$~19~\kms\ (\citealt{Grier13b}). However, for completeness, we reproduce Figure 23 from P14 in our Figure~\ref{fig:m-sigma} with our two additional measurements included. We see that the locations of both PG\,2130+099 and 3C\,120 are also consistent with the distribution of dynamical black hole mass measurements.

All four of our targets exhibit kinematics with either mostly near-circular elliptical
orbits or near-circular elliptical orbits combined with inflowing gas. In addition, all of these kinematic
results are consistent with the qualitative interpretation of the MEMECHO analysis we performed in Paper
I; the velocity-delay maps we recovered all show signatures of
elliptical motion and/or inflow. Our results are consistent with all
of the targets favoring dynamics that are dominated by the
gravitational potential of the black hole, supporting the use of
reverberation mapping to measure $M_{\rm BH}$. 

Reverberation mapping data for three of our targets (3C\,120, Mrk\,335,
and PG\,2130+099) have also recently been analyzed by \cite{Li13}, who
use a geometry-only model of the BLR within the framework that is based on the model proposed by \citet{Pancoast11}.
The main difference between this work and the analysis of \cite{Li13}
is that we model the \Hbeta\ emission line-profile while \cite{Li13}
model the integrated \Hbeta\ flux.  We also use slightly different
models for the BLR geometry, with \cite{Li13} including non-linear
response and this analysis including additional asymmetry parameters
for the geometry ($\kappa$, $\gamma$, and $\xi$). While formally consistent, we find smaller inclination angles (between 25-30 degrees rather than closer to 50 degrees) and smaller opening angles (also ranging from about 15-30 degrees rather than closer to 50) for 3C\,120 and Mrk\,335. The measurements for PG\,2130+099 differ significantly from those reported by \cite{Li13}, likely due to the differences in modeling approaches. The BLR radii reported by \cite{Li13} for the three targets are formally 
consistent with this those found in this work, though our models constrain these angles much more tightly; this highlights the importance of using full dynamical models when data of sufficient quality are available. 

\subsection{Systematic Uncertainties in the Model} 
\label{modelcaveats} 

The formal uncertainties from our model fitting are very small (particularly in the case of Mrk\,335): With this particular data set, we are approaching the regime where the uncertainties in the inferred parameters are dominated by the systematic limitations of the model rather than the observational signal-to-noise ratio or cadence of the data.  Because our formal (statistical) uncertainties are derived from the data, it is possible that the true uncertainties are somewhat larger than the statistical uncertainties alone. 
We now discuss possible sources of systematic uncertainty from both correlations between model parameters and from physical assumptions made by the model.  

\subsubsection{Systematic Uncertainties from Correlations between Model Parameters} 
\label{sec:angles} 
There are two primary ways that correlations between the model parameters influence the inferred uncertainties in BLR properties.  
First, constraints provided by the data may introduce degeneracies in parameter space that appear unphysical when observed for a sample of AGNs as a whole.  The best example of this for the current sample of 9 AGNs with BLR modeling is the correlation between the inclination and opening angles.  As shown in Figure~\ref{fig:2dposteriors}, the inclination and opening angles show a positive tight correlation and approximately equal values.  Inspecting emission-line profiles produced by the BLR model with similar parameters as inferred for these data suggests that the opening angle is forced to be at least as large as the inclination angle in order to generate a single-peaked emission-line profile.  As the opening angle becomes larger than the inclination angle, the transfer function becomes more spread out in wavelength and time-lag space. If the data prefer a more compact transfer function, it will thus force the opening angle to be as small as possible while still producing a single-peaked line profile. This condition is met when the opening angle is very close to the inclination angle. Since there is no obvious upper bound to possible values of the opening angle, it is likely that the values we infer are closer to upper limits; in other words, to match a single-peaked emission-line profile the effective prior on the opening angle is from $\theta_i \to 90$ deg, so an inference of $\theta_o \sim \theta_i$ could also mean that $\theta_o \lesssim \theta_i$.  
While it is impossible to quantify the magnitude of this systematic uncertainty without comparing to models that include other methods for making single-peaked emission lines, we can test whether the inclination or opening angle has a greater effect on the model transfer function. We varied $\theta_i$ and $\theta_o$ by 5 to 30 degrees around a fiducial value of $\theta_i = \theta_o = 30$ deg and compared the resulting transfer functions.  We found that the inclination angle has a qualitatively greater effect by causing more extreme changes in the line profile shape. Comparing the relative dispersion in differences of the transfer function to that derived from the fiducial parameters confirms this result quantitatively. This suggests that the inclination angle may be more robustly determined than the opening angle; and indeed, the consistency between the radio jet inclination measurement for 3C\,120 and our model (see Section~\ref{sec:threec}) supports the idea that our modeling more robustly constrains the inclination angle. 

Second, with such a flexibly parameterized model there can be multiple distinct parameter combinations that end up producing the same distribution of point particles in position and velocity space.  A clear example of this is for inferred models where the inclination and opening angles approach 90 degrees with $\gamma \sim 5$, as for Mrk~335, Mrk~1501, and PG\,2130+099.  This combination of parameter values corresponds to a spherical distribution of point particles, but where the particles are concentrated along the face-on axis that is perpendicular to the observer's line of sight, such that they form a jet-like structure. Since rotations in the plane of the sky cannot be resolved by RM, which is only sensitive to time delay and line-of-sight velocity, this jet perpendicular to the line of sight is equivalent to a face-on thick disk. This may increase the statistical uncertainty in certain inferred parameters compared to the true uncertainty in, for example, disk thickness.  However, Figures~\ref{fig:mrk335_posteriors} --\ref{fig:pg2130_posteriors} show that the number of posterior samples in the solutions with larger inclination and opening angles just described is a small fraction of the total and our use of the median instead of the mean value of the posterior PDF in Table~\ref{Table:tbl2} minimizes the contribution from posterior samples in the tails of the distribution.

\subsubsection{Systematic Uncertainties from Assumptions of the Model} 
In order to develop a simply parameterized and flexible model for the BLR, many assumptions were made about BLR physics, including the following: 
\begin{itemize} 
\item We assume spatially and temporally uniform responsivity across the BLR without optical depth effects. However, \cite{Korista04} show that spatially and temporally constant responsivity is not necessarily a good assumption for Balmer lines. 
\item We assume that the only non-negligible force at play is gravity; thus any force that has a functional form of 1/r$^2$ (such as radiation pressure from electron scattering from a distant source) is subsumed under this and we cannot differentiate between them. 
\item We assume that the driving continuum light curve is emitted by a source much smaller than the BLR --- in our case, a point-like source at $r$ = 0 (i.e., we are neglecting to account for the size of the ionizing continuum-emitting region).
\end{itemize} 

To fully address the first point above would require a self-consistent BLR model that includes photoionization models, constraining both the emission properties and gas distribution simultaneously. We note, however, that other BLR models (e.g., MEMECHO) and codes to measure time lags (e.g., {\tt JAVELIN}) currently all include a linear response of the emission lines, working under the assumption that the changes in AGN luminosity within the time spanned by a single RM campaign are small enough that the response will not deviate far from linearity. To address the second assumption above would require an updated model for the inner accretion disk and the emission of ionizing photons, which is quite challenging to constrain. As such, for these two assumptions, we are unable to calculate or estimate the possible magnitude of systematic uncertainty introduced in our measurements. 

However, in light of recent developments, the assumption that the driving optical continuum originates at $r = 0$ warrants further discussion. As part of a multiwavelength RM campaign using the Hubble Space Telescope, \cite{Fausnaugh16a} measure a time delay between the optical continuum at various wavelengths and the ultraviolet (UV) continuum at 1367 \AA. In particular, they found that the $V$-band continuum lags the  UV continuum by about 2 days, indicating that the $V$-band emitting region is at least 2 light-days farther out than the UV-emitting region. This is on par with the mean lag of $\sim$ 2.5 days measured in the optical and UV \heii \ emission lines --- the size of the optical continuum-emitting region is therefore far from negligible. It is possible that our assumption of continuum emission originating at $r = 0$ could cause us to underestimate $M_{\rm BH}$, although it is also worth noting that NGC\,5548 was in an exceptional state during the course of this RM campaign and the emission-line time delays were all shorter than predicted from the radius-luminosity relation (\citealt{Pei16}). Time delays between the EUV and optical of varying quality and significance have been reported by several other studies as well (\citealt{Collier98}; \citealt{Sergeev05}; \citealt{Mchardy14}; \citealt{Shappee14}; \citealt{Edelson15}), suggesting that the situation is not unique. 

For NGC\,5548, \cite{Pei16} determine that the \Hbeta-UV lag is about two days longer than the \Hbeta-optical lag (consistent with the optical-to-UV lag reported by \citealt{Fausnaugh16a}), so the BLR radius is underestimated by about 50\% in this particular case. The magnitude of the effect on \mbh caused by the non-negligible accretion disk size will depend on the AGN, as the accretion disk size depends on luminosity, \mbh, and the slope of the temperature profile. However, \citealt{Pei16} examine this in detail and find that the scaling with luminosity is expected to be slow and the scatter in the radius-luminosity relationship (\citealt{Bentz13}) is small --- thus this effect is likely small for most AGN. Additional studies to measure the relationship between the optical and UV lags would be useful to confirm this for sources of different \mbh and luminosities. 

It is worth noting that this particular systematic will not have an effect on BH masses measured via traditional RM (by measuring the average time lag within the BLR via cross-correlation or some other method), as the use of the \msigma relation to calculate the average scale factor $f$ automatically makes up for this by requiring that AGN fall on the quiescent \msigma relation. For our sample of four AGN, the scale factors measured from the dynamical modeling are consistent with the average scale factor measured from the \msigma relation (with the exception of PG\,2130+099), which also suggests that any systematic effect caused by neglecting the size of the continuum-emitting region is \mbox{likely to be small in these targets.} 

\subsection{The Scale Factor $f$}

The scale factor $f$ summarizes the relationship between the observables from an RM experiment ($\Delta V$ and $R$) and $M_{\rm BH}$. 
The line width and BLR radius are related as $\Delta V \propto r^{-1/2}$, indicating that the quantity $\Delta V^2$$R$/G, or the virial product, is proportional 
to \mbh --- however, there are other quantities that affect \mbh measurements that are not measured in traditional RM. 
For example, if the BLR is a disk or disk-like, $f$ should depend strongly on the inclination of the disk relative to the line of sight. In addition, 
it should depend on the kinematics of the BLR, which could include infall, outflow, or circular-like motion. The scale factor $f$ should also depend on both the radial distribution of gas within the BLR and its responsivity.  It will depend on how well the characteristic $R$ measured by RM actually reflects the typical size of the system. If the BLR environment is dependent on accretion rate, one might
expect to see a correlation between $f$ and the AGN luminosity and/or
the Eddington ratio. We do not expect $f$ to correlate with \mbh --- such a correlation would indicate that the BLR structure and kinematics are mass-dependent or likely that we have some sort of selection bias in our sample. 

As discussed in Section \ref{sec:introduction}, traditional RM studies do not yield measurements of $f$; the most common way to 
estimate $f$ to obtain \mbh measurements using RM is
by assuming that AGNs follow the same \msigma relationship as
quiescent galaxies to calculate $\bar{f}$ for the entire set of
AGNs. There has been a range of many different $\bar{f}$
measurements reported in the literature, ranging from 
 $\rm{log}_{10}$$\bar{f_{\sigma}}$~=~0.58 (\citealt{Graham11}) to 0.74 (\citealt{Onken04}), though most are consistent with one another to within the uncertainties. It is also possible that galaxies with different morphologies follow different \msigma relations, which would require the use of different values of $f$ depending on the galaxy type (e.g., \citealt{Ho14}). However, by measuring $f$ in individual AGN using dynamical modeling, we can consider the values of $f$ for our sample as a whole, as described in Sections \ref{sec:indiv_f} and \ref{sec:fcorrelations}. 

\subsubsection{$\bar{f}$ Measurements} 
\label{sec:indiv_f}

We calculate the mean scale factor ($\bar{f}$) in our sample to compare with the
external measurements reported in previous studies. 
To obtain a measurement of $\bar{f}$ from a sample of
sources with dynamical modeling results, we model the distribution of
posterior PDFs of individual $f$ values using a Gaussian with a mean
$f$ and dispersion or scatter in individual values of $f$ (see
P14 for further details).  The posterior PDFs of individual $f$ values are calculated by dividing the posterior PDF of black hole
mass by the virial product for each target.  This analysis yields
uncertainties on both the mean scale factor $\bar{f}$ and the dispersion or
scatter.  For the AGN10 sample, we calculate the mean value of
$\log_{10}f_\sigma$ to be $\rm{log}_{10}$$\bar{f_{\sigma}}$~=~\avgfmysample \ and the dispersion in $\log_{10}f_\sigma$
to be \dispavgfmysample.  We also calculate the mean value of $\log_{10}f_{\rm FWHM}$
to be $\rm{log}_{10}$$\bar{f}_{\rm FWHM}$~=~\avgffwhmmysample \ and the dispersion in $\log_{10}f_{\rm
  FWHM}$ to be \dispavgffwhmmysample. Our  $\rm{log}_{10}$$\bar{f_{\sigma}}$ is consistent to within the uncertainties 
with the measurement made by P14 (who report a mean value in
$\log_{10}f_\sigma$ of  $\rm{log}_{10}$$\bar{f_{\sigma}}$~=~
0.68 $\pm$ 0.40 and a dispersion in $\log_{10}f_\sigma$ of
0.75~$\pm$~0.40), and our value for $\rm{log}_{10}$$\bar{f}_{\rm FWHM}$ is also consistent with their
measurement of \mbox{$\rm{log}_{10}$$\bar{f}_{\rm FWHM}~=-0.07~
\pm$~0.40} and dispersion of 0.77~$\pm$~0.38. 

\begin{figure}
\begin{center}
\includegraphics[scale = 0.4, angle = 0, trim = 0 0 0 0, clip]{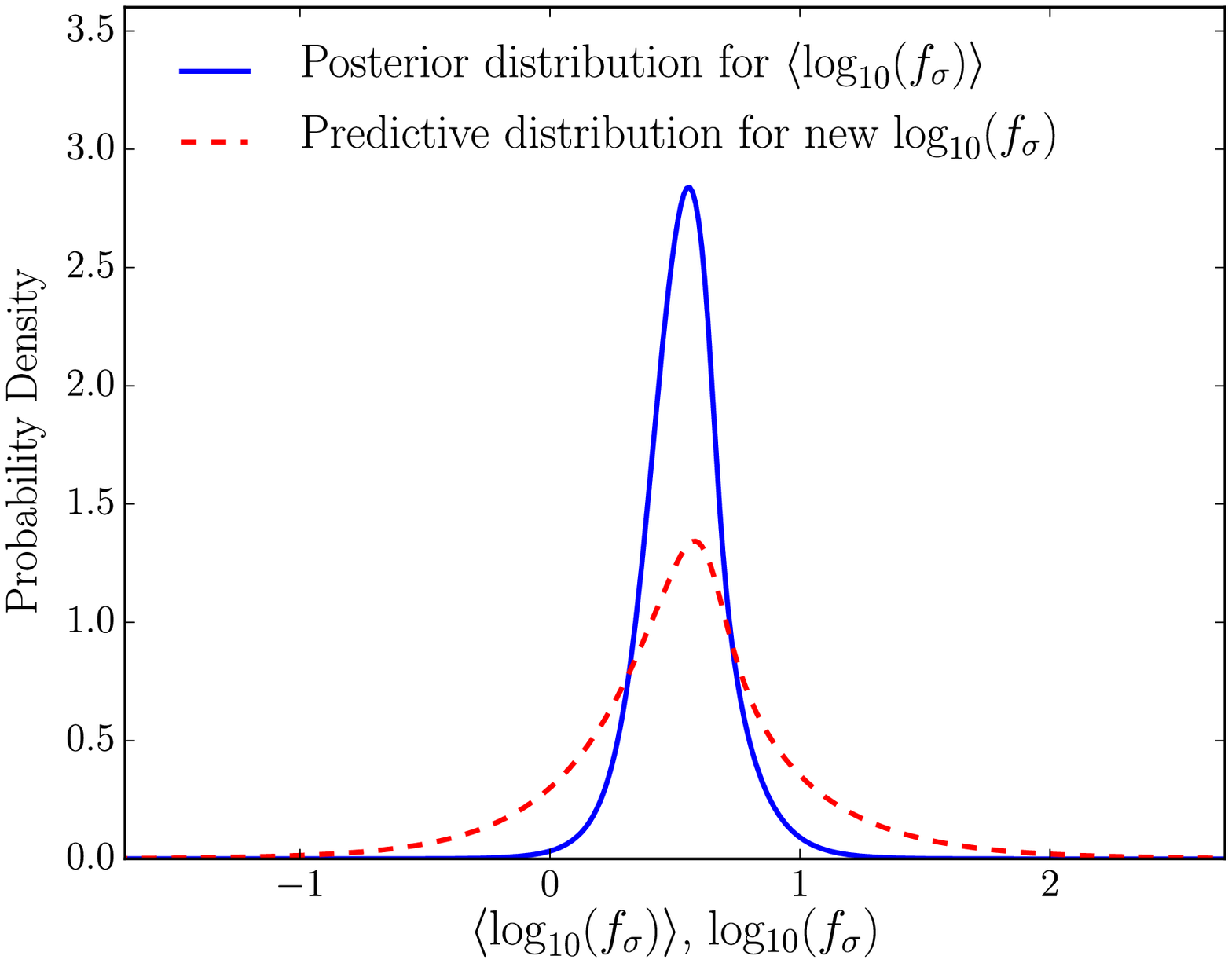}
\includegraphics[scale = 0.4, angle = 0, trim = 0 0 0 0, clip]{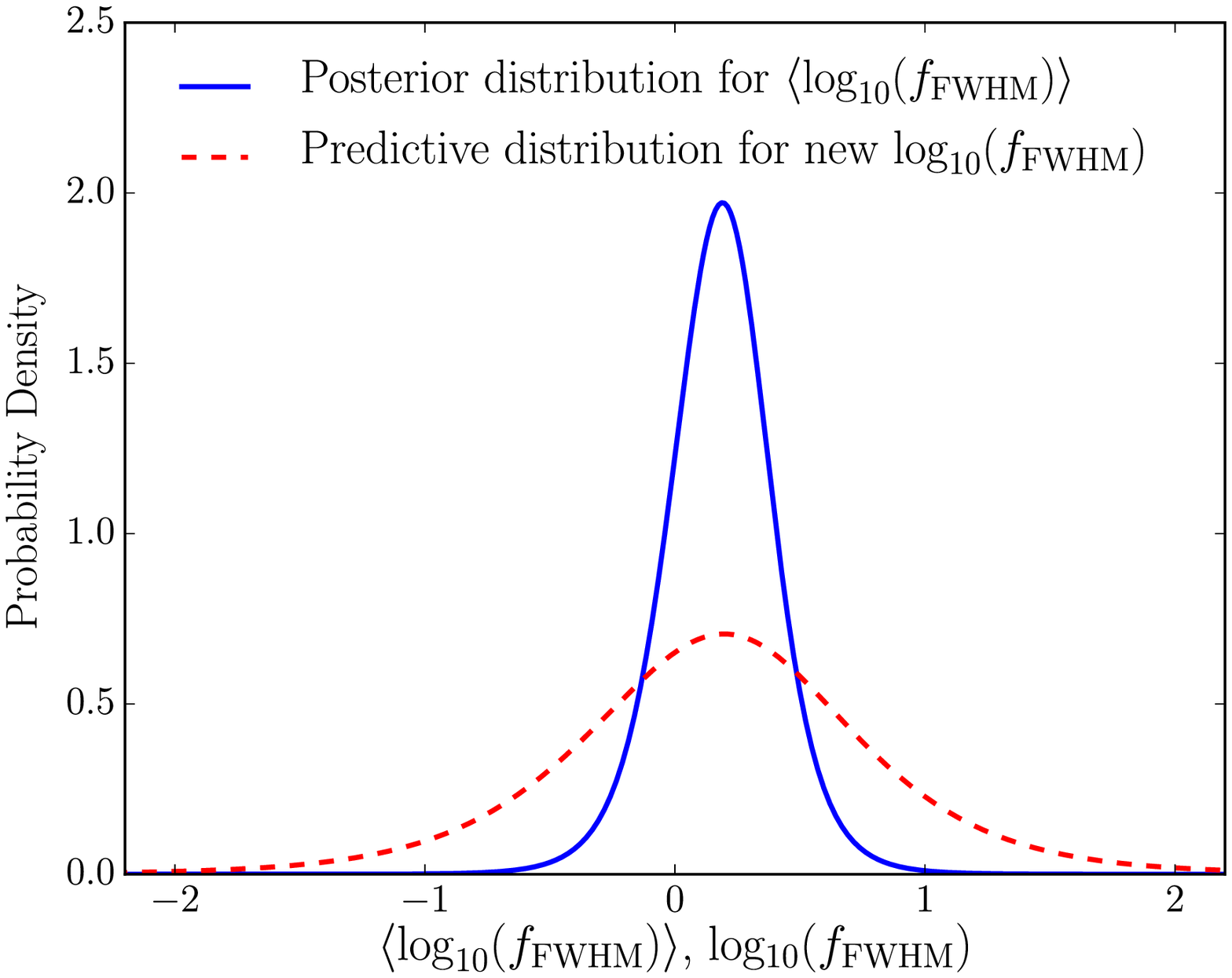}
\caption{Top panel: The posterior distribution of the mean of
  log$_{10}(f_{\sigma})$ for the combined sample of AGN10 and P14
  AGNs. The posterior for the mean of the $f_{\sigma}$ distribution is shown in blue,
  and the predictive distribution for new measurements of $f_{\sigma}$ is shown as the
  dashed red line. Bottom panel: The posterior distribution of the mean
  of log$_{10}(f_{\rm FWHM})$ for the combined sample of AGN10 and P14
  AGNs. The posterior for the mean of the $f_{\rm FWHM}$ distribution is shown in blue,
  and the predictive distribution for new measurements of $f_{\rm FWHM}$ is shown as the
  dashed red line. }
\label{fig:mean-f}
\end{center}
\end{figure}

When we combine our sample with the P14 sample for a total of 9 AGNs
(hereafter referred to as the ``combined sample''), we measure
a mean value of $\log_{10}f_\sigma$ of $\rm{log}_{10}$$\bar{f_{\sigma}}$~=~\avgfwholesample \ with a dispersion in $\log_{10}f_\sigma$
of \dispavgfwholesample \ and $\rm{log}_{10}$$\bar{f}_{\rm FWHM}$~=~\avgffwhmwholesample \ with a dispersion in
$\log_{10}f_{\rm FWHM}$ of \dispavgffwhmwholesample.  We show the
posterior and predictive distributions for $ \log_{10}
f_{\sigma}$ and $\rm{log}_{10}$$f_{\rm FWHM}$
in Figure~\ref{fig:mean-f}.  The predictive distributions are the
distributions from which new measurements of $f_{\sigma}$ or $f_{\rm
  FWHM}$ are drawn and are generated from linear combinations of Gaussians weighted by the
posterior probability of the model parameters fit to the 9 measured values of $f$.  
The $\bar{f_{\sigma}}$ value is consistent with the
measurements of $\bar{f_{\sigma}}$ found using the \msigma
relation listed above, suggesting that the \msigma relation yields a reasonable calibration for AGN \mbh measurements. 

\begin{figure*}
\begin{center}
\includegraphics[scale = 0.55, angle = -90, trim = 0 0 0 0, clip]{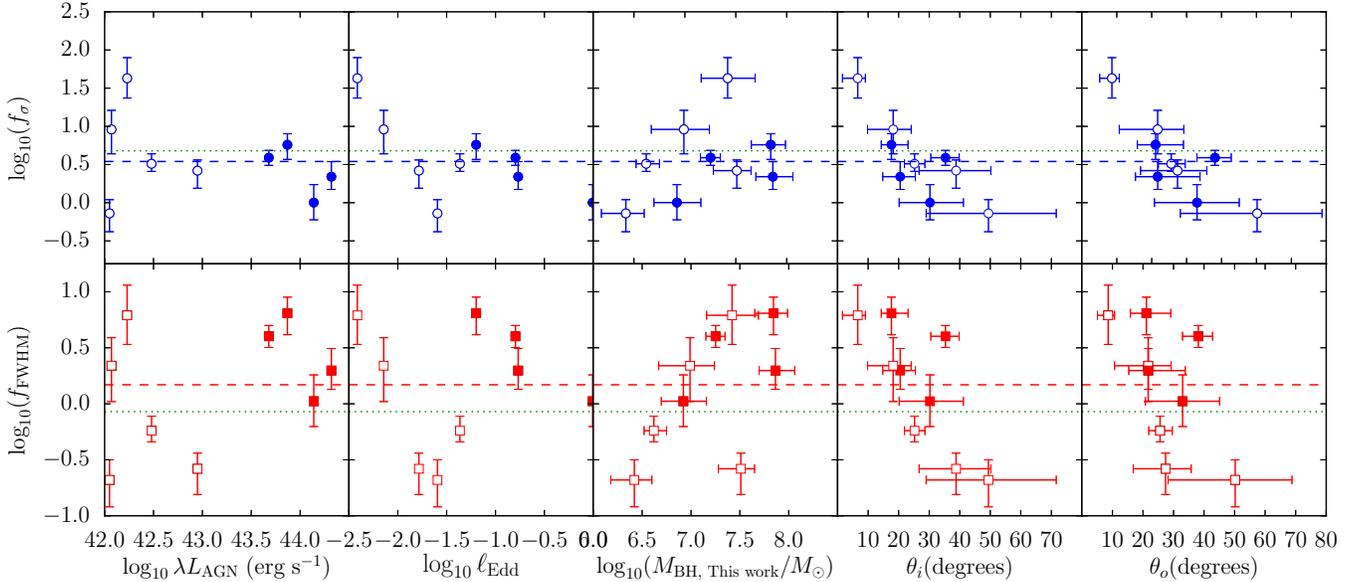}
\caption{The quantities log$_{10}(f_{\rm \sigma})$ (top panels) and log$_{10}(f_{\rm
    FHWM})$ (bottom panels) versus various AGN properties (either measured independently or inferred from the modeling). Measurements for our four targets are shown by filled circles and squares, and the five targets from P14 are shown by open circles and squares. The blue and red dashed lines show the
  average log$_{10}(\bar{f_{\rm \sigma}})$ and log$_{10}(\bar{f}_{\rm FHWM})$
  (red) of the entire sample of nine targets, and the green dotted
  lines show the mean values reported by P14 for their sample of five.}
\label{fig:all-fplots}
\end{center}
\end{figure*}

\subsubsection{Correlations between $f$ and AGN Properties}
\label{sec:fcorrelations} 
RM is a very time- and observation-intensive practice; the data
used for modeling in this work and by P14 have very high
signal-to-noise ratio and high cadence. RM \mbh measurements, determinations of the scale factor $f$, and BLR information are impractical to obtain in large
samples of AGNs due to the stringent data quality requirements. With our expansion of the sample of AGNs with
dynamical modeling results, we not only learn about the BLR in individual
sources, but also aim to uncover any potential correlations between
$f$ and other more easily-measured AGN parameters that may offer
information without the heavy requirements of RM campaigns and enable improvements in single-epoch \mbh measurements. To search
for such correlations, we combine the AGN10 sample with the P14
sample, nearly doubling the number of sources with detailed constraints
on the BLR. However, there is one other AGN, Mrk\,50, which has dynamical modeling results
(\citealt{Pancoast12}). The results for this source were obtained
using a less flexible BLR model that did not allow for
unbound inflowing and outflowing orbits. The narrow line model has
also been updated, and more asymmetry parameters ($\gamma$ and $\xi$)
have been added to account for the complexities of the BLR geometry. Since  it is
difficult to directly compare the results for Mrk\,50 to our new
results and those of P14, we omit this source from our analysis.

We searched for correlations between $f$ and all of the model parameters 
reported in Table~\ref{Table:tbl2}; however, only a few correlations were found. 
In Figure~\ref{fig:all-fplots} we show the scale factors
measured as a function of five different parameters: $L_{\rm AGN}$, Eddington ratio ($\ell_{\rm Edd} = L/L_{\rm Edd}$), $M_{\rm BH}$ recovered from dynamical
modeling, $\theta_i$, and $\theta_o$. Spearman rank test results between $f$ and these parameters are given in Table~\ref{Table:tbl4}. 
Luminosities ($L_{5100}$) were measured by G12 and $\ell_{\rm Edd}$ values were computed using these luminosities, a bolometric correction factor of 9, and the \mbh measurements reported by our dynamical modeling. The luminosities have all been corrected for
host galaxy contamination by \cite{Bentz13}, with the exception of Mrk\,1501, which 
does not yet have imaging data of sufficient quality to make these measurements. The host galaxy contribution to the luminosity of Mrk\,1501 causes the AGN luminosity to be 
overestimated. While our data do not span a sufficient wavelength range to make precise measurements of the host contamination, we use our spectral decomposition of the mean spectrum to estimate that the stellar component contributes roughly 20\% of the total flux at 5100 \AA. This is comparable to that seen in the other three objects that have more well-measured host components (see G12). Applying this correction would change the AGN luminosity of Mrk\,1501 from log$\lambda L_{\rm AGN}~=~44.32$ to log$\lambda L_{\rm AGN}$~=~44.22, which has no significant effect on the position of Mrk\,1501 in Figure~\ref{fig:all-fplots}. Since this estimate is so uncertain and has no effect on the correlation, we choose to use the uncorrected value in this work. We note that the AGN10 sample occupies a higher range of
AGN luminosities than the P14 sample, as all five P14 targets have
log$_{10}L_{\rm AGN} < 43.0$ while all four AGN10 targets have
log$_{10}L_{\rm AGN} > 43.0$. 

 \begin{deluxetable}{lcrc} 
\tablewidth{0pt} 
\tablecaption{Spearman Correlation Test Results} 
\tablehead{ 
\colhead{Parameter 1} & 
\colhead{Parameter 2} &  
\colhead{$\rho$}  & 
\colhead{p-value}  
 } 
\startdata 
log$_{10}$($f_{\sigma}$)   & 	log$_{10}\lambda L_{\rm AGN}$ & 	$ -0.217 $  & 0.576  \\ 
log$_{10}$($f_{\rm FWHM}$) & 	log$_{10}\lambda L_{\rm AGN}$ & 	$  0.267 $  & 0.488  \\ 
log$_{10}$($f_{\sigma}$)   & 	log$_{10}(\ell_{\rm Edd}$) 	& 	$ -0.550 $  & 0.125  \\ 
log$_{10}$($f_{\rm FWHM}$) & 	log$_{10}(\ell_{\rm Edd}$)  	& 	$ -0.017 $  & 0.966  \\ 
log$_{10}$($f_{\sigma}$)   & 	log$_{10}(M_{\rm BH}$) 		& 	$  0.317 $  & 0.406  \\ 
log$_{10}$($f_{\rm FWHM}$) & 	log$_{10}(M_{\rm BH}$)  		& 	$  0.533 $  & 0.139  \\ 
log$_{10}$($f_{\sigma}$)   & 	$\theta_i$ 					& 	$ -0.767 $  & 0.016  \\ 
log$_{10}$($f_{\rm FWHM}$) &	$\theta_i$					    & 	$ -0.800 $  & 0.010  \\ 
log$_{10}$($f_{\sigma}$)   & 	$\theta_o$ 				    & 	$ -0.767 $  & 0.016  \\ 
log$_{10}$($f_{\rm FWHM}$) & 	$\theta_o$ 				    &	$ -0.683 $  & 0.042      
\enddata  
\tablecaption{Spearman Rank correlation coefficient $\rho$ for the combined AGN10+ P14 sample.}   
\label{Table:tbl4}  
\end{deluxetable}   


Figure~\ref{fig:all-fplots} shows no significant correlation between $f$ and the 5100\,\AA \ luminosity of the
AGNs in either the AGN10 sample or the combined sample. 
However, there is a possible correlation between $f_{\sigma}$ and $\ell_{\rm Edd}$, though it is not seen with $f_{\rm FWHM}$; such a correlation may be expected if radiation pressure forces provided an important contribution. However, we note that interpretation of the $\ell_{\rm Edd}$ panel is not entirely straightforward, as both of the parameters being examined ($f$ and $\ell_{\rm Edd}$) are computed using \mbh measured by our model (as a reminder, $f$ is calculated by dividing \mbh measured by the dynamical modeling by the virial product measured from the line width and mean BLR radius, and $\ell_{\rm Edd}$ is also calculated using \mbh from the dynamical modeling). 

Figure~\ref{fig:all-fplots} also suggests a possible correlation between $f$ and $M_{\rm BH}$, particularly for $f_{\rm FWHM}$. This correlation would be expected only if the BLR geometry or dynamics depends on the size of the BH itself. 
We also see a correlation between $f$ and inclination angle and also between $f$ and opening angle, and these
correlations are strengthened when the two samples are combined. P14 also observed the correlation between $f$ and the inclination angle
of the system in their sample, noting that this correlation was
predicted by \cite{Goad12} but could also be an effect of correlated
errors. We see strong correlations between \mbh and inclination angle (and between inclination and opening angles; see Section \ref{sec:angles} and Figure \ref{fig:2dposteriors}), so the correlations seen between $f$ and these parameters could be related to this. 

We have a broad range in line profile shape among
the combined sample, so we also investigated potential correlations
between the line shape, parameterized by the ratio of FWHM/$\sigma$ (with FWHM measured from the mean spectrum and $\sigma$ measured from the RMS residual spectrum), and various BLR quantities recovered by the
models such as inclination angle, $f$, and $M_{\rm BH}$. The \Hbeta \ profiles in the AGN10 sample have similar FWHM/$\sigma$ that are much lower than those of the P14 sample. We see no evidence for any correlation between BLR parameters and log$_{10}$(FWHM/$\sigma$) among the AGN10 sample
or the combined sample --- the tentative correlation seen by P14
between log$_{10}$(FWHM/$\sigma$) and \mbh disappears when the combined
sample is examined.

\section{SUMMARY}
\label{sec:summary}
We have used the dynamical modeling techniques of \cite{Pancoast14b} to
constrain the geometry and kinematics of the \Hbeta-emitting BLR in
a sample of four AGNs from a 2010 RM campaign. The main results of
this work are as follows.
\begin{enumerate}
\item In all cases, we find that the \Hbeta-emitting BLR is
  best-described by a thick, inclined disk that is closer to face-on
  than edge-on relative to the observer.
\item In the case of 3C\,120, our measurements of the inclination angle are consistent with measurements made independently using the radio jet inclination. This indicates that the BLR and jet axes are aligned in this system, and demonstrates the potential combined power of dynamical modeling and radio jet measurements in the future (see Section \ref{sec:threec}). 
\item As with the P14 sample, our results are consistent with most of the broad line 
  emission originating from the far side of the BLR. This is consistent with photoionization modeling (e.g., \citealt{Ferland92}; \citealt{Obrien94}; \citealt{Korista04}). 
\item We see signatures of gas in near-circular elliptical orbits as well as inflowing
  movement of the gas within the BLR in all four cases. This
  is in good agreement with the signatures seen in the velocity-delay
  maps of these sources in Paper I. 
  \item From our recovered models, we obtain \mbh measurements for all
  four targets: log$_{10}$($M_{\rm BH}$)~=~\mrkAmbh \ for Mrk\,335,
  log$_{10}$($M_{\rm BH}$)~=~\mrkBmbh \ for Mrk\,1501,
  log$_{10}$($M_{\rm BH}$)~=~\threecmbh \ for 3C\,120, and
  log$_{10}$($M_{\rm BH}$)~=~\pgmbh \ for PG\,2130+099. These
  measurements are independent of $f$ and largely consistent with
  previous RM \mbh measurements, with the exception of PG\,2130+099. This lack of agreement with previous measurements could be because PG\,2130+099 has a different scale factor $f$, or because its behavior is not well-represented within the constraints of this particular model. 
\item We compute the scale factors ($f$) for all four objects using virial products measured by G12 and \mbh measurements from our models
  (tabulated in Table~\ref{Table:tbl3}). Three of the four have values close to the average value typically used in RM \mbh calculations; however, PG\,2130+099 has a much lower scale factor $f$ than the others. We note that previous studies of PG\,2130+099 (e.g., \citealt{Grier08}) have had difficulties measuring a reliable lag for this source; the pattern of variability in this object has always led to ambiguous results. 
\item We find that $f$ is possibly correlated with $M_{\rm BH}$, $\ell_{\rm Edd}$, inclination
  angle, and opening angle of the system, but not any other parameters that we
  examined. We similarly see no correlations between the ratio of the
  FWHM to the line dispersion of the \Hbeta \ emission line and any of
  the following: $M_{\rm BH}$, inclination angle, or $f$.
\item We combine the posterior distributions of $f$ for each AGN in
  our sample and measure the mean scale factor $\bar{f}$ for our sample of four
  AGNs to be  $\rm{log}_{10}$$\bar{f_{\sigma}}$~=~\avgfmysample . We measure this for 
  the combined sample from this study, which includes objects from P14, to be $\rm{log}_{10}$$\bar{f_{\sigma}}$~=~\avgfwholesample. Our $\bar{f}$ measurements are consistent with nearly all measurements of $\bar{f}$ made using the \msigma relationship. 

\end{enumerate} 
 
The dynamical modeling method from P14 was strikingly successful in providing constraints for these four AGNs, likely due to the high S/N and high cadence of the RM campaign, allowing us to nearly double the size of the sample of objects having been modeled in this manner.  Though we have expanded the size of the sample with dynamical modeling measurements, our sample is still small and not entirely representative of the reverberation-mapped population of AGNs, much less the AGN population as a whole. Thus, while promising, it may still be premature to use our average measured scale factor $\bar{f}$ to calibrate the \mbh scale. Future RM experiments with similar data quality would yield additional constraints on the larger population of AGNs by allowing us to improve the statistical significance of observed correlations between various parameters, and eventually allow us to build up a large enough sample to apply the average scale factor $f$ and its scatter to the broader AGN sample with confidence. 

 \acknowledgments CJG and BMP acknowledge support from NSF AST-1008882. AP is supported by NASA through Einstein Postdoctoral Fellowship grant number PF5-160141 awarded by the Chandra X-ray Center. Research by AJB is supported by NSF grant AST-1412693. MMF acknowledges support from NSF grant AST-1008882 and a fellowship awarded by the Ohio State University. BJB used resources provided by the Centre for eResearch at the University of Auckland. TT acknowledges support from the Packard Foundation in the form of a Packard Fellowship, and by the National Science Foundation through grant AST-1412315. 


\begin{thebibliography}{}
\expandafter\ifx\csname natexlab\endcsname\relax\def\natexlab#1{#1}\fi

\bibitem[{{Agudo} {et~al.}(2012){Agudo}, {G{\'o}mez}, {Casadio}, {Cawthorne},
  \& {Roca-Sogorb}}]{Agudo12}
{Agudo}, I., {G{\'o}mez}, J.~L., {Casadio}, C., {Cawthorne}, T.~V., \&
  {Roca-Sogorb}, M. 2012, \apj, 752, 92

\bibitem[{{Barth} {et~al.}(2011{\natexlab{a}}){Barth}, {Nguyen}, {Malkan},
  {Filippenko}, {Li}, {Gorjian}, {Joner}, {Bennert}, {Botyanszki}, {Cenko},
  {Childress}, {Choi}, {Comerford}, {Cucciara}, {da Silva}, {Duch{\^e}ne},
  {Fumagalli}, {Ganeshalingam}, {Gates}, {Gerke}, {Griffith}, {Harris},
  {Hintz}, {Hsiao}, {Kandrashoff}, {Keel}, {Kirkman}, {Kleiser}, {Laney},
  {Lee}, {Lopez}, {Lowe}, {Moody}, {Morton}, {Nierenberg}, {Nugent},
  {Pancoast}, {Rex}, {Rich}, {Silverman}, {Smith}, {Sonnenfeld}, {Suzuki},
  {Tytler}, {Walsh}, {Woo}, {Yang}, \& {Zeisse}}]{Barth11b}
{Barth}, A.~J., {Nguyen}, M.~L., {Malkan}, M.~A., {et~al.} 2011{\natexlab{a}},
  \apj, 732, 121

\bibitem[{{Barth} {et~al.}(2011{\natexlab{b}}){Barth}, {Pancoast}, {Thorman},
  {Bennert}, {Sand}, {Li}, {Canalizo}, {Filippenko}, {Gates}, {Greene},
  {Malkan}, {Stern}, {Treu}, {Woo}, {Assef}, {Bae}, {Brewer}, {Buehler},
  {Cenko}, {Clubb}, {Cooper}, {Diamond-Stanic}, {Hiner}, {H{\"o}nig}, {Joner},
  {Kandrashoff}, {Laney}, {Lazarova}, {Nierenberg}, {Park}, {Silverman}, {Son},
  {Sonnenfeld}, {Tollerud}, {Walsh}, {Walters}, {da Silva}, {Fumagalli},
  {Gregg}, {Harris}, {Hsiao}, {Lee}, {Lopez}, {Rex}, {Suzuki}, {Trump},
  {Tytler}, {Worseck}, \& {Yesuf}}]{Barth11}
{Barth}, A.~J., {Pancoast}, A., {Thorman}, S.~J., {et~al.} 2011{\natexlab{b}},
  \apjl, 743, L4

\bibitem[{{Barth} {et~al.}(2013){Barth}, {Pancoast}, {Bennert}, {Brewer},
  {Canalizo}, {Filippenko}, {Gates}, {Greene}, {Li}, {Malkan}, {Sand}, {Stern},
  {Treu}, {Woo}, {Assef}, {Bae}, {Buehler}, {Cenko}, {Clubb}, {Cooper},
  {Diamond-Stanic}, {H{\"o}nig}, {Joner}, {Laney}, {Lazarova}, {Nierenberg},
  {Silverman}, {Tollerud}, \& {Walsh}}]{Barth13}
{Barth}, A.~J., {Pancoast}, A., {Bennert}, V.~N., {et~al.} 2013, \apj, 769, 128

\bibitem[{{Barth} {et~al.}(2015){Barth}, {Bennert}, {Canalizo}, {Filippenko},
  {Gates}, {Greene}, {Li}, {Malkan}, {Pancoast}, {Sand}, {Stern}, {Treu},
  {Woo}, {Assef}, {Bae}, {Brewer}, {Cenko}, {Clubb}, {Cooper},
  {Diamond-Stanic}, {Hiner}, {H{\"o}nig}, {Hsiao}, {Kandrashoff}, {Lazarova},
  {Nierenberg}, {Rex}, {Silverman}, {Tollerud}, \& {Walsh}}]{Barth15}
{Barth}, A.~J., {Bennert}, V.~N., {Canalizo}, G., {et~al.} 2015, \apjs, 217, 26

\bibitem[{{Batiste} {et~al.}(2016){Batiste}, {Bentz}, {Raimundo},
  {Vestergaard}, \& {Onken}}]{Batiste17}
{Batiste}, M., {Bentz}, M.~C., {Raimundo}, S.~I., {Vestergaard}, M., \&
  {Onken}, C.~A. 2016, ArXiv e-prints, arXiv:1612.02815

\bibitem[{{Bentz} {et~al.}(2009){Bentz}, {Peterson}, {Netzer}, {Pogge}, \&
  {Vestergaard}}]{Bentz09a}
{Bentz}, M.~C., {Peterson}, B.~M., {Netzer}, H., {Pogge}, R.~W., \&
  {Vestergaard}, M. 2009, \apj, 697, 160

\bibitem[{{Bentz} {et~al.}(2010){Bentz}, {Walsh}, {Barth}, {Yoshii}, {Woo},
  {Wang}, {Treu}, {Thornton}, {Street}, {Steele}, {Silverman}, {Serduke},
  {Sakata}, {Minezaki}, {Malkan}, {Li}, {Lee}, {Hiner}, {Hidas}, {Greene},
  {Gates}, {Ganeshalingam}, {Filippenko}, {Canalizo}, {Bennert}, \&
  {Baliber}}]{Bentz10a}
{Bentz}, M.~C., {Walsh}, J.~L., {Barth}, A.~J., {et~al.} 2010, \apj, 716, 993

\bibitem[{{Bentz} {et~al.}(2013){Bentz}, {Denney}, {Grier}, {Barth},
  {Peterson}, {Vestergaard}, {Bennert}, {Canalizo}, {De Rosa}, {Filippenko},
  {Gates}, {Greene}, {Li}, {Malkan}, {Pogge}, {Stern}, {Treu}, \&
  {Woo}}]{Bentz13}
{Bentz}, M.~C., {Denney}, K.~D., {Grier}, C.~J., {et~al.} 2013, \apj, 767, 149

\bibitem[{{Blandford} \& {McKee}(1982)}]{Blandford82}
{Blandford}, R.~D., \& {McKee}, C.~F. 1982, \apj, 255, 419

\bibitem[{{Boroson} \& {Green}(1992)}]{Boroson92}
{Boroson}, T.~A., \& {Green}, R.~F. 1992, \apjs, 80, 109

\bibitem[{{Brewer} {et~al.}(2010){Brewer}, {P{\'a}rtay}, \&
  {Cs{\'a}nyi}}]{Brewer10}
{Brewer}, B.~J., {P{\'a}rtay}, L.~B., \& {Cs{\'a}nyi}, G. 2010, ascl:1010.029

\bibitem[{{Brewer} {et~al.}(2011){Brewer}, {Treu}, {Pancoast}, {Barth},
  {Bennert}, {Bentz}, {Filippenko}, {Greene}, {Malkan}, \& {Woo}}]{Brewer11}
{Brewer}, B.~J., {Treu}, T., {Pancoast}, A., {et~al.} 2011, \apjl, 733, L33

\bibitem[{{Collier} {et~al.}(1998){Collier}, {Horne}, {Kaspi}, {Netzer},
  {Peterson}, {Wanders}, {Alexander}, {Bertram}, {Comastri}, {Gaskell},
  {Malkov}, {Maoz}, {Mignoli}, {Pogge}, {Pronik}, {Sergeev}, {Snedden},
  {Stirpe}, {Bochkarev}, {Burenkov}, {Shapovalova}, \& {Wagner}}]{Collier98}
{Collier}, S.~J., {Horne}, K., {Kaspi}, S., {et~al.} 1998, \apj, 500, 162

\bibitem[{{De Rosa} {et~al.}(2015){De Rosa}, {Peterson}, {Ely}, {Kriss},
  {Crenshaw}, {Horne}, {Korista}, {Netzer}, {Pogge}, {Ar{\'e}valo}, {Barth},
  {Bentz}, {Brandt}, {Breeveld}, {Brewer}, {Dalla Bont{\`a}}, {De
  Lorenzo-C{\'a}ceres}, {Denney}, {Dietrich}, {Edelson}, {Evans}, {Fausnaugh},
  {Gehrels}, {Gelbord}, {Goad}, {Grier}, {Grupe}, {Hall}, {Kaastra}, {Kelly},
  {Kennea}, {Kochanek}, {Lira}, {Mathur}, {McHardy}, {Nousek}, {Pancoast},
  {Papadakis}, {Pei}, {Schimoia}, {Siegel}, {Starkey}, {Treu}, {Uttley},
  {Vaughan}, {Vestergaard}, {Villforth}, {Yan}, {Young}, \& {Zu}}]{Derosa15}
{De Rosa}, G., {Peterson}, B.~M., {Ely}, J., {et~al.} 2015, \apj, 806, 128

\bibitem[{{Denney} {et~al.}(2009){Denney}, {Peterson}, {Pogge}, {Adair},
  {Atlee}, {Au-Yong}, {Bentz}, {Bird}, {Brokofsky}, {Chisholm}, {Comins},
  {Dietrich}, {Doroshenko}, {Eastman}, {Efimov}, {Ewald}, {Ferbey}, {Gaskell},
  {Hedrick}, {Jackson}, {Klimanov}, {Klimek}, {Kruse}, {Lad{\'e}route}, {Lamb},
  {Leighly}, {Minezaki}, {Nazarov}, {Onken}, {Petersen}, {Peterson},
  {Poindexter}, {Sakata}, {Schlesinger}, {Sergeev}, {Skolski}, {Stieglitz},
  {Tobin}, {Unterborn}, {Vestergaard}, {Watkins}, {Watson}, \&
  {Yoshii}}]{Denney09c}
{Denney}, K.~D., {Peterson}, B.~M., {Pogge}, R.~W., {et~al.} 2009, \apjl, 704,
  L80

\bibitem[{{Denney} {et~al.}(2010){Denney}, {Peterson}, {Pogge}, {Adair},
  {Atlee}, {Au-Yong}, {Bentz}, {Bird}, {Brokofsky}, {Chisholm}, {Comins},
  {Dietrich}, {Doroshenko}, {Eastman}, {Efimov}, {Ewald}, {Ferbey}, {Gaskell},
  {Hedrick}, {Jackson}, {Klimanov}, {Klimek}, {Kruse}, {Lad{\'e}route}, {Lamb},
  {Leighly}, {Minezaki}, {Nazarov}, {Onken}, {Petersen}, {Peterson},
  {Poindexter}, {Sakata}, {Schlesinger}, {Sergeev}, {Skolski}, {Stieglitz},
  {Tobin}, {Unterborn}, {Vestergaard}, {Watkins}, {Watson}, \&
  {Yoshii}}]{Denney10}
---. 2010, \apj, 721, 715

\bibitem[{{Doroshenko} {et~al.}(2012){Doroshenko}, {Sergeev}, {Klimanov},
  {Pronik}, \& {Efimov}}]{Doroshenko12}
{Doroshenko}, V.~T., {Sergeev}, S.~G., {Klimanov}, S.~A., {Pronik}, V.~I., \&
  {Efimov}, Y.~S. 2012, \mnras, 426, 416

\bibitem[{{Du} {et~al.}(2016){Du}, {Lu}, {Hu}, {Qiu}, {Li}, {Huang}, {Wang},
  {Bai}, {Bian}, {Yuan}, {Ho}, {Wang}, \& {SEAMBH Collaboration}}]{Du16}
{Du}, P., {Lu}, K.-X., {Hu}, C., {et~al.} 2016, \apj, 820, 27

\bibitem[{{Edelson} {et~al.}(2015){Edelson}, {Gelbord}, {Horne}, {McHardy},
  {Peterson}, {Ar{\'e}valo}, {Breeveld}, {De Rosa}, {Evans}, {Goad}, {Kriss},
  {Brandt}, {Gehrels}, {Grupe}, {Kennea}, {Kochanek}, {Nousek}, {Papadakis},
  {Siegel}, {Starkey}, {Uttley}, {Vaughan}, {Young}, {Barth}, {Bentz},
  {Brewer}, {Crenshaw}, {Dalla Bont{\`a}}, {De Lorenzo-C{\'a}ceres}, {Denney},
  {Dietrich}, {Ely}, {Fausnaugh}, {Grier}, {Hall}, {Kaastra}, {Kelly},
  {Korista}, {Lira}, {Mathur}, {Netzer}, {Pancoast}, {Pei}, {Pogge},
  {Schimoia}, {Treu}, {Vestergaard}, {Villforth}, {Yan}, \& {Zu}}]{Edelson15}
{Edelson}, R., {Gelbord}, J.~M., {Horne}, K., {et~al.} 2015, \apj, 806, 129

\bibitem[{{Fausnaugh} {et~al.}(2016){Fausnaugh}, {Denney}, {Barth}, {Bentz},
  {Bottorff}, {Carini}, {Croxall}, {De Rosa}, {Goad}, {Horne}, {Joner},
  {Kaspi}, {Kim}, {Klimanov}, {Kochanek}, {Leonard}, {Netzer}, {Peterson},
  {Schn{\"u}lle}, {Sergeev}, {Vestergaard}, {Zheng}, {Zu}, {Anderson},
  {Ar{\'e}valo}, {Bazhaw}, {Borman}, {Boroson}, {Brandt}, {Breeveld}, {Brewer},
  {Cackett}, {Crenshaw}, {Dalla Bont{\`a}}, {De Lorenzo-C{\'a}ceres},
  {Dietrich}, {Edelson}, {Efimova}, {Ely}, {Evans}, {Filippenko}, {Flatland},
  {Gehrels}, {Geier}, {Gelbord}, {Gonzalez}, {Gorjian}, {Grier}, {Grupe},
  {Hall}, {Hicks}, {Horenstein}, {Hutchison}, {Im}, {Jensen}, {Jones},
  {Kaastra}, {Kelly}, {Kennea}, {Kim}, {Korista}, {Kriss}, {Lee}, {Lira},
  {MacInnis}, {Manne-Nicholas}, {Mathur}, {McHardy}, {Montouri}, {Musso},
  {Nazarov}, {Norris}, {Nousek}, {Okhmat}, {Pancoast}, {Papadakis}, {Parks},
  {Pei}, {Pogge}, {Pott}, {Rafter}, {Rix}, {Saylor}, {Schimoia}, {Siegel},
  {Spencer}, {Starkey}, {Sung}, {Teems}, {Treu}, {Turner}, {Uttley},
  {Villforth}, {Weiss}, {Woo}, {Yan}, \& {Young}}]{Fausnaugh16a}
{Fausnaugh}, M.~M., {Denney}, K.~D., {Barth}, A.~J., {et~al.} 2016, \apj, 821,
  56

\bibitem[{{Ferland} {et~al.}(1992){Ferland}, {Peterson}, {Horne}, {Welsh}, \&
  {Nahar}}]{Ferland92}
{Ferland}, G.~J., {Peterson}, B.~M., {Horne}, K., {Welsh}, W.~F., \& {Nahar},
  S.~N. 1992, \apj, 387, 95

\bibitem[{{Goad} {et~al.}(2012){Goad}, {Korista}, \& {Ruff}}]{Goad12}
{Goad}, M.~R., {Korista}, K.~T., \& {Ruff}, A.~J. 2012, \mnras, 426, 3086

\bibitem[{{Goad} {et~al.}(2016){Goad}, {Korista}, {De Rosa}, {Kriss},
  {Edelson}, {Barth}, {Ferland}, {Kochanek}, {Netzer}, {Peterson}, {Bentz},
  {Bisogni}, {Crenshaw}, {Denney}, {Ely}, {Fausnaugh}, {Grier}, {Gupta},
  {Horne}, {Kaastra}, {Pancoast}, {Pei}, {Pogge}, {Skielboe}, {Starkey},
  {Vestergaard}, {Zu}, {Anderson}, {Ar{\'e}valo}, {Bazhaw}, {Borman},
  {Boroson}, {Bottorff}, {Brandt}, {Breeveld}, {Brewer}, {Cackett}, {Carini},
  {Croxall}, {Dalla Bont{\`a}}, {De Lorenzo-C{\'a}ceres}, {Dietrich},
  {Efimova}, {Evans}, {Filippenko}, {Flatland}, {Gehrels}, {Geier}, {Gelbord},
  {Gonzalez}, {Gorjian}, {Grupe}, {Hall}, {Hicks}, {Horenstein}, {Hutchison},
  {Im}, {Jensen}, {Joner}, {Jones}, {Kaspi}, {Kelly}, {Kennea}, {Kim}, {Kim},
  {Klimanov}, {Lee}, {Leonard}, {Lira}, {MacInnis}, {Manne-Nicholas}, {Mathur},
  {McHardy}, {Montouri}, {Musso}, {Nazarov}, {Norris}, {Nousek}, {Okhmat},
  {Papadakis}, {Parks}, {Pott}, {Rafter}, {Rix}, {Saylor}, {Schimoia},
  {Schn{\"u}lle}, {Sergeev}, {Siegel}, {Spencer}, {Sung}, {Teems}, {Treu},
  {Turner}, {Uttley}, {Villforth}, {Weiss}, {Woo}, {Yan}, {Young}, \&
  {Zheng}}]{Goad16}
{Goad}, M.~R., {Korista}, K.~T., {De Rosa}, G., {et~al.} 2016, \apj, 824, 11

\bibitem[{{Graham} {et~al.}(2011){Graham}, {Onken}, {Athanassoula}, \&
  {Combes}}]{Graham11}
{Graham}, A.~W., {Onken}, C.~A., {Athanassoula}, E., \& {Combes}, F. 2011,
  \mnras, 412, 2211

\bibitem[{{Grier} {et~al.}(2008){Grier}, {Peterson}, {Bentz}, {Denney},
  {Eastman}, {Dietrich}, {Pogge}, {Prieto}, {DePoy}, {Assef}, {Atlee}, {Bird},
  {Eyler}, {Peeples}, {Siverd}, {Watson}, \& {Yee}}]{Grier08}
{Grier}, C.~J., {Peterson}, B.~M., {Bentz}, M.~C., {et~al.} 2008, \apj, 688,
  837

\bibitem[{{Grier} {et~al.}(2012{\natexlab{a}}){Grier}, {Peterson}, {Pogge},
  {Denney}, {Bentz}, {Martini}, {Sergeev}, {Kaspi}, {Zu}, {Kochanek},
  {Shappee}, {Stanek}, {Araya Salvo}, {Beatty}, {Bird}, {Bord}, {Borman},
  {Che}, {Chen}, {Cohen}, {Dietrich}, {Doroshenko}, {Efimov}, {Free},
  {Ginsburg}, {Henderson}, {Horne}, {King}, {Mogren}, {Molina}, {Mosquera},
  {Nazarov}, {Okhmat}, {Pejcha}, {Rafter}, {Shields}, {Skowron}, {Szczygiel},
  {Valluri}, \& {van Saders}}]{Grier12a}
{Grier}, C.~J., {Peterson}, B.~M., {Pogge}, R.~W., {et~al.} 2012{\natexlab{a}},
  \apjl, 744, L4

\bibitem[{{Grier} {et~al.}(2012{\natexlab{b}}){Grier}, {Peterson}, {Pogge},
  {Denney}, {Bentz}, {Martini}, {Sergeev}, {Kaspi}, {Minezaki}, {Zu},
  {Kochanek}, {Siverd}, {Shappee}, {Stanek}, {Araya Salvo}, {Beatty}, {Bird},
  {Bord}, {Borman}, {Che}, {Chen}, {Cohen}, {Dietrich}, {Doroshenko}, {Drake},
  {Efimov}, {Free}, {Ginsburg}, {Henderson}, {King}, {Koshida}, {Mogren},
  {Molina}, {Mosquera}, {Nazarov}, {Okhmat}, {Pejcha}, {Rafter}, {Shields},
  {Skowron}, {Szczygiel}, {Valluri}, \& {van Saders}}]{Grier12b}
---. 2012{\natexlab{b}}, \apj, 755, 60 (G12)

\bibitem[{{Grier} {et~al.}(2013{\natexlab{a}}){Grier}, {Martini}, {Watson},
  {Peterson}, {Bentz}, {Dasyra}, {Dietrich}, {Ferrarese}, {Pogge}, \&
  {Zu}}]{Grier13b}
{Grier}, C.~J., {Martini}, P., {Watson}, L.~C., {et~al.} 2013{\natexlab{a}},
  \apj, 773, 90

\bibitem[{{Grier} {et~al.}(2013{\natexlab{b}}){Grier}, {Peterson}, {Horne},
  {Bentz}, {Pogge}, {Denney}, {De Rosa}, {Martini}, {Kochanek}, {Zu},
  {Shappee}, {Siverd}, {Beatty}, {Sergeev}, {Kaspi}, {Araya Salvo}, {Bird},
  {Bord}, {Borman}, {Che}, {Chen}, {Cohen}, {Dietrich}, {Doroshenko}, {Efimov},
  {Free}, {Ginsburg}, {Henderson}, {King}, {Mogren}, {Molina}, {Mosquera},
  {Nazarov}, {Okhmat}, {Pejcha}, {Rafter}, {Shields}, {Skowron}, {Szczygiel},
  {Valluri}, \& {van Saders}}]{Grier13a}
{Grier}, C.~J., {Peterson}, B.~M., {Horne}, K., {et~al.} 2013{\natexlab{b}},
  \apj, 764, 47 (Paper I)

\bibitem[{{Ho} \& {Kim}(2014)}]{Ho14}
{Ho}, L.~C., \& {Kim}, M. 2014, \apj, 789, 17

\bibitem[{{Horne}(1994)}]{Horne94}
{Horne}, K. 1994, in Astronomical Society of the Pacific Conference Series,
  Vol.~69, Reverberation Mapping of the Broad-Line Region in Active Galactic
  Nuclei, ed. P.~M. {Gondhalekar}, K.~{Horne}, \& B.~M. {Peterson}, 23--25

\bibitem[{{Horne} {et~al.}(1991){Horne}, {Welsh}, \& {Peterson}}]{Horne91}
{Horne}, K., {Welsh}, W.~F., \& {Peterson}, B.~M. 1991, \apjl, 367, L5

\bibitem[{{Hu} {et~al.}(2015){Hu}, {Du}, {Lu}, {Li}, {Wang}, {Qiu}, {Bai},
  {Kaspi}, {Ho}, {Netzer}, {Wang}, \& {SEAMBH Collaboration}}]{Hu15}
{Hu}, C., {Du}, P., {Lu}, K.-X., {et~al.} 2015, \apj, 804, 138

\bibitem[{{Jorstad} {et~al.}(2005){Jorstad}, {Marscher}, {Lister}, {Stirling},
  {Cawthorne}, {Gear}, {G{\'o}mez}, {Stevens}, {Smith}, {Forster}, \&
  {Robson}}]{Jorstad05}
{Jorstad}, S.~G., {Marscher}, A.~P., {Lister}, M.~L., {et~al.} 2005, \aj, 130,
  1418

\bibitem[{{Kaspi} {et~al.}(2000){Kaspi}, {Smith}, {Netzer}, {Maoz}, {Jannuzi},
  \& {Giveon}}]{Kaspi00}
{Kaspi}, S., {Smith}, P.~S., {Netzer}, H., {et~al.} 2000, \apj, 533, 631

\bibitem[{{Kassebaum} {et~al.}(1997){Kassebaum}, {Peterson}, {Wanders},
  {Pogge}, {Bertram}, \& {Wagner}}]{Kassebaum97}
{Kassebaum}, T.~M., {Peterson}, B.~M., {Wanders}, I., {et~al.} 1997, \apj, 475,
  106

\bibitem[{{Kelly} {et~al.}(2009){Kelly}, {Bechtold}, \&
  {Siemiginowska}}]{Kelly09}
{Kelly}, B.~C., {Bechtold}, J., \& {Siemiginowska}, A. 2009, \apj, 698, 895

\bibitem[{{Kollatschny} {et~al.}(2014){Kollatschny}, {Ulbrich}, {Zetzl},
  {Kaspi}, \& {Haas}}]{Kollatschny14}
{Kollatschny}, W., {Ulbrich}, K., {Zetzl}, M., {Kaspi}, S., \& {Haas}, M. 2014,
  ArXiv e-prints, arXiv:1405.1588

\bibitem[{{Korista} \& {Goad}(2004)}]{Korista04}
{Korista}, K.~T., \& {Goad}, M.~R. 2004, \apj, 606, 749

\bibitem[{{Kova{\v c}evi{\'c}} {et~al.}(2010){Kova{\v c}evi{\'c}},
  {Popovi{\'c}}, \& {Dimitrijevi{\'c}}}]{Kovacevic10}
{Kova{\v c}evi{\'c}}, J., {Popovi{\'c}}, L.~{\v C}., \& {Dimitrijevi{\'c}},
  M.~S. 2010, \apjs, 189, 15

\bibitem[{{Koz{\l}owski}(2016)}]{Kozlowski16}
{Koz{\l}owski}, S. 2016, \apj, 826, 118

\bibitem[{{Krolik} \& {Done}(1995)}]{Krolik95}
{Krolik}, J.~H., \& {Done}, C. 1995, \apj, 440, 166

\bibitem[{{Li} {et~al.}(2013){Li}, {Wang}, {Ho}, {Du}, \& {Bai}}]{Li13}
{Li}, Y.-R., {Wang}, J.-M., {Ho}, L.~C., {Du}, P., \& {Bai}, J.-M. 2013, \apj,
  779, 110

\bibitem[{{MacLeod} {et~al.}(2010){MacLeod}, {Ivezi{\'c}}, {Kochanek},
  {Koz{\l}owski}, {Kelly}, {Bullock}, {Kimball}, {Sesar}, {Westman}, {Brooks},
  {Gibson}, {Becker}, \& {de Vries}}]{MacLeod10}
{MacLeod}, C.~L., {Ivezi{\'c}}, {\v Z}., {Kochanek}, C.~S., {et~al.} 2010,
  \apj, 721, 1014

\bibitem[{{Marscher} {et~al.}(2002){Marscher}, {Jorstad}, {G{\'o}mez}, {Aller},
  {Ter{\"a}sranta}, {Lister}, \& {Stirling}}]{Marscher02}
{Marscher}, A.~P., {Jorstad}, S.~G., {G{\'o}mez}, J.-L., {et~al.} 2002, \nat,
  417, 625

\bibitem[{{McConnell} \& {Ma}(2013)}]{McConnell13}
{McConnell}, N.~J., \& {Ma}, C.-P. 2013, \apj, 764, 184

\bibitem[{{McHardy} {et~al.}(2014){McHardy}, {Cameron}, {Dwelly}, {Connolly},
  {Lira}, {Emmanoulopoulos}, {Gelbord}, {Breedt}, {Arevalo}, \&
  {Uttley}}]{Mchardy14}
{McHardy}, I.~M., {Cameron}, D.~T., {Dwelly}, T., {et~al.} 2014, \mnras, 444,
  1469

\bibitem[{{Nelson} \& {Whittle}(1995)}]{Nelson95}
{Nelson}, C.~H., \& {Whittle}, M. 1995, \apjs, 99, 67

\bibitem[{{O'Brien} {et~al.}(1994){O'Brien}, {Goad}, \&
  {Gondhalekar}}]{Obrien94}
{O'Brien}, P.~T., {Goad}, M.~R., \& {Gondhalekar}, P.~M. 1994, \mnras, 268, 845

\bibitem[{{Onken} {et~al.}(2004){Onken}, {Ferrarese}, {Merritt}, {Peterson},
  {Pogge}, {Vestergaard}, \& {Wandel}}]{Onken04}
{Onken}, C.~A., {Ferrarese}, L., {Merritt}, D., {et~al.} 2004, \apj, 615, 645

\bibitem[{{Pancoast} {et~al.}(2011){Pancoast}, {Brewer}, \&
  {Treu}}]{Pancoast11}
{Pancoast}, A., {Brewer}, B.~J., \& {Treu}, T. 2011, \apj, 730, 139

\bibitem[{{Pancoast} {et~al.}(2014{\natexlab{a}}){Pancoast}, {Brewer}, \&
  {Treu}}]{Pancoast14b}
---. 2014{\natexlab{a}}, \mnras, 445, 3055

\bibitem[{{Pancoast} {et~al.}(2014{\natexlab{b}}){Pancoast}, {Brewer}, {Treu},
  {Park}, {Barth}, {Bentz}, \& {Woo}}]{Pancoast14}
{Pancoast}, A., {Brewer}, B.~J., {Treu}, T., {et~al.} 2014{\natexlab{b}},
  \mnras, 445, 3073

\bibitem[{{Pancoast} {et~al.}(2012){Pancoast}, {Brewer}, {Treu}, {Barth},
  {Bennert}, {Canalizo}, {Filippenko}, {Gates}, {Greene}, {Li}, {Malkan},
  {Sand}, {Stern}, {Woo}, {Assef}, {Bae}, {Buehler}, {Cenko}, {Clubb},
  {Cooper}, {Diamond-Stanic}, {Hiner}, {H{\"o}nig}, {Joner}, {Kandrashoff},
  {Laney}, {Lazarova}, {Nierenberg}, {Park}, {Silverman}, {Son}, {Sonnenfeld},
  {Thorman}, {Tollerud}, {Walsh}, \& {Walters}}]{Pancoast12}
---. 2012, \apj, 754, 49

\bibitem[{{Park} {et~al.}(2012){Park}, {Woo}, {Treu}, {Barth}, {Bentz},
  {Bennert}, {Canalizo}, {Filippenko}, {Gates}, {Greene}, {Malkan}, \&
  {Walsh}}]{Park12}
{Park}, D., {Woo}, J.-H., {Treu}, T., {et~al.} 2012, \apj, 747, 30

\bibitem[{{Pei} {et~al.}(2017){Pei}, {Fausnaugh}, {Barth}, {Peterson}, {Bentz},
  {De Rosa}, {Denney}, {Goad}, {Kochanek}, {Korista}, {Kriss}, {Pogge},
  {Bennert}, {Brotherton}, {Clubb}, {Dalla Bont{\`a}}, {Filippenko}, {Greene},
  {Grier}, {Vestergaard}, {Zheng}, {Adams}, {Beatty}, {Bigley}, {Brown},
  {Brown}, {Canalizo}, {Comerford}, {Coker}, {Corsini}, {Croft}, {Croxall},
  {Deason}, {Eracleous}, {Fox}, {Gates}, {Henderson}, {Holmbeck}, {Holoien},
  {Jensen}, {Johnson}, {Kelly}, {Kim}, {King}, {Lau}, {Li}, {Lochhaas}, {Ma},
  {Manne-Nicholas}, {Mauerhan}, {Malkan}, {McGurk}, {Morelli}, {Mosquera},
  {Mudd}, {Muller Sanchez}, {Nguyen}, {Ochner}, {Ou-Yang}, {Pancoast}, {Penny},
  {Pizzella}, {Poleski}, {Runnoe}, {Scott}, {Schimoia}, {Shappee}, {Shivvers},
  {Simonian}, {Siviero}, {Somers}, {Stevens}, {Strauss}, {Tayar}, {Tejos},
  {Treu}, {Van Saders}, {Vican}, {Villanueva}, {Yuk}, {Zakamska}, {Zhu},
  {Anderson}, {Ar{\'e}valo}, {Bazhaw}, {Bisogni}, {Borman}, {Bottorff},
  {Brandt}, {Breeveld}, {Cackett}, {Carini}, {Crenshaw}, {De
  Lorenzo-C{\'a}ceres}, {Dietrich}, {Edelson}, {Efimova}, {Ely}, {Evans},
  {Ferland}, {Flatland}, {Gehrels}, {Geier}, {Gelbord}, {Grupe}, {Gupta},
  {Hall}, {Hicks}, {Horenstein}, {Horne}, {Hutchison}, {Im}, {Joner}, {Jones},
  {Kaastra}, {Kaspi}, {Kelly}, {Kennea}, {Kim}, {Kim}, {Klimanov}, {Lee},
  {Leonard}, {Lira}, {MacInnis}, {Mathur}, {McHardy}, {Montouri}, {Musso},
  {Nazarov}, {Netzer}, {Norris}, {Nousek}, {Okhmat}, {Papadakis}, {Parks},
  {Pott}, {Rafter}, {Rix}, {Saylor}, {Schn{\"u}lle}, {Sergeev}, {Siegel},
  {Skielboe}, {Spencer}, {Starkey}, {Sung}, {Teems}, {Turner}, {Uttley},
  {Villforth}, {Weiss}, {Woo}, {Yan}, {Young}, \& {Zu}}]{Pei16}
{Pei}, L., {Fausnaugh}, M.~M., {Barth}, A.~J., {et~al.} 2017, ArXiv e-prints,
  arXiv:1702.01177

\bibitem[{{Peterson}(1993)}]{Peterson93}
{Peterson}, B.~M. 1993, \pasp, 105, 247

\bibitem[{{Peterson} {et~al.}(1998){Peterson}, {Wanders}, {Bertram}, {Hunley},
  {Pogge}, \& {Wagner}}]{Peterson98}
{Peterson}, B.~M., {Wanders}, I., {Bertram}, R., {et~al.} 1998, \apj, 501, 82

\bibitem[{{Peterson} {et~al.}(2004){Peterson}, {Ferrarese}, {Gilbert}, {Kaspi},
  {Malkan}, {Maoz}, {Merritt}, {Netzer}, {Onken}, {Pogge}, {Vestergaard}, \&
  {Wandel}}]{Peterson04}
{Peterson}, B.~M., {Ferrarese}, L., {Gilbert}, K.~M., {et~al.} 2004, \apj, 613,
  682

\bibitem[{{Schlegel} {et~al.}(1998){Schlegel}, {Finkbeiner}, \&
  {Davis}}]{Schlegel98}
{Schlegel}, D.~J., {Finkbeiner}, D.~P., \& {Davis}, M. 1998, \apj, 500, 525

\bibitem[{{Sergeev} {et~al.}(2005){Sergeev}, {Doroshenko}, {Golubinskiy},
  {Merkulova}, \& {Sergeeva}}]{Sergeev05}
{Sergeev}, S.~G., {Doroshenko}, V.~T., {Golubinskiy}, Y.~V., {Merkulova},
  N.~I., \& {Sergeeva}, E.~A. 2005, \apj, 622, 129

\bibitem[{{Shappee} {et~al.}(2014){Shappee}, {Prieto}, {Grupe}, {Kochanek},
  {Stanek}, {De Rosa}, {Mathur}, {Zu}, {Peterson}, {Pogge}, {Komossa}, {Im},
  {Jencson}, {Holoien}, {Basu}, {Beacom}, {Szczygie{\l}}, {Brimacombe},
  {Adams}, {Campillay}, {Choi}, {Contreras}, {Dietrich}, {Dubberley},
  {Elphick}, {Foale}, {Giustini}, {Gonzalez}, {Hawkins}, {Howell}, {Hsiao},
  {Koss}, {Leighly}, {Morrell}, {Mudd}, {Mullins}, {Nugent}, {Parrent},
  {Phillips}, {Pojmanski}, {Rosing}, {Ross}, {Sand}, {Terndrup}, {Valenti},
  {Walker}, \& {Yoon}}]{Shappee14}
{Shappee}, B.~J., {Prieto}, J.~L., {Grupe}, D., {et~al.} 2014, \apj, 788, 48

\bibitem[{{Skielboe} {et~al.}(2015){Skielboe}, {Pancoast}, {Treu}, {Park},
  {Barth}, \& {Bentz}}]{Skielboe15}
{Skielboe}, A., {Pancoast}, A., {Treu}, T., {et~al.} 2015, \mnras, 454, 144

\bibitem[{{van Groningen} \& {Wanders}(1992)}]{vanGroningen92}
{van Groningen}, E., \& {Wanders}, I. 1992, \pasp, 104, 700

\bibitem[{{V{\'e}ron-Cetty} {et~al.}(2004){V{\'e}ron-Cetty}, {Joly}, \&
  {V{\'e}ron}}]{Veroncetty04}
{V{\'e}ron-Cetty}, M.-P., {Joly}, M., \& {V{\'e}ron}, P. 2004, \aap, 417, 515

\bibitem[{{Walsh} {et~al.}(2009){Walsh}, {Minezaki}, {Bentz}, {Barth},
  {Baliber}, {Li}, {Stern}, {Bennert}, {Brown}, {Canalizo}, {Filippenko},
  {Gates}, {Greene}, {Malkan}, {Sakata}, {Street}, {Treu}, {Woo}, \&
  {Yoshii}}]{Walsh09}
{Walsh}, J.~L., {Minezaki}, T., {Bentz}, M.~C., {et~al.} 2009, \apjs, 185, 156

\bibitem[{{Wills} \& {Browne}(1986)}]{Wills86}
{Wills}, B.~J., \& {Browne}, I.~W.~A. 1986, \apj, 302, 56

\bibitem[{{Woo} {et~al.}(2015){Woo}, {Yoon}, {Park}, {Park}, \& {Kim}}]{Woo15}
{Woo}, J.-H., {Yoon}, Y., {Park}, S., {Park}, D., \& {Kim}, S.~C. 2015, \apj,
  801, 38

\bibitem[{{Zu} {et~al.}(2013){Zu}, {Kochanek}, {Koz{\l}owski}, \&
  {Udalski}}]{Zu13}
{Zu}, Y., {Kochanek}, C.~S., {Koz{\l}owski}, S., \& {Udalski}, A. 2013, \apj,
  765, 106

\bibitem[{{Zu} {et~al.}(2011){Zu}, {Kochanek}, \& {Peterson}}]{Zu11}
{Zu}, Y., {Kochanek}, C.~S., \& {Peterson}, B.~M. 2011, \apj, 735, 80

\end{thebibliography}

\appendix  

We here present the posterior distributions for some of the key model parameters for each target. Each parameter is defined in Section~\ref{sec:modelingmethods}. We also show two-dimensional posterior distributions to demonstrate the correlations between $M_{\rm BH}$, $\theta_i$, and $\theta_o$. 
\begin{figure*}
\begin{center}
\includegraphics[scale = 0.45, angle = 0, trim = 0 0 0 0, clip]{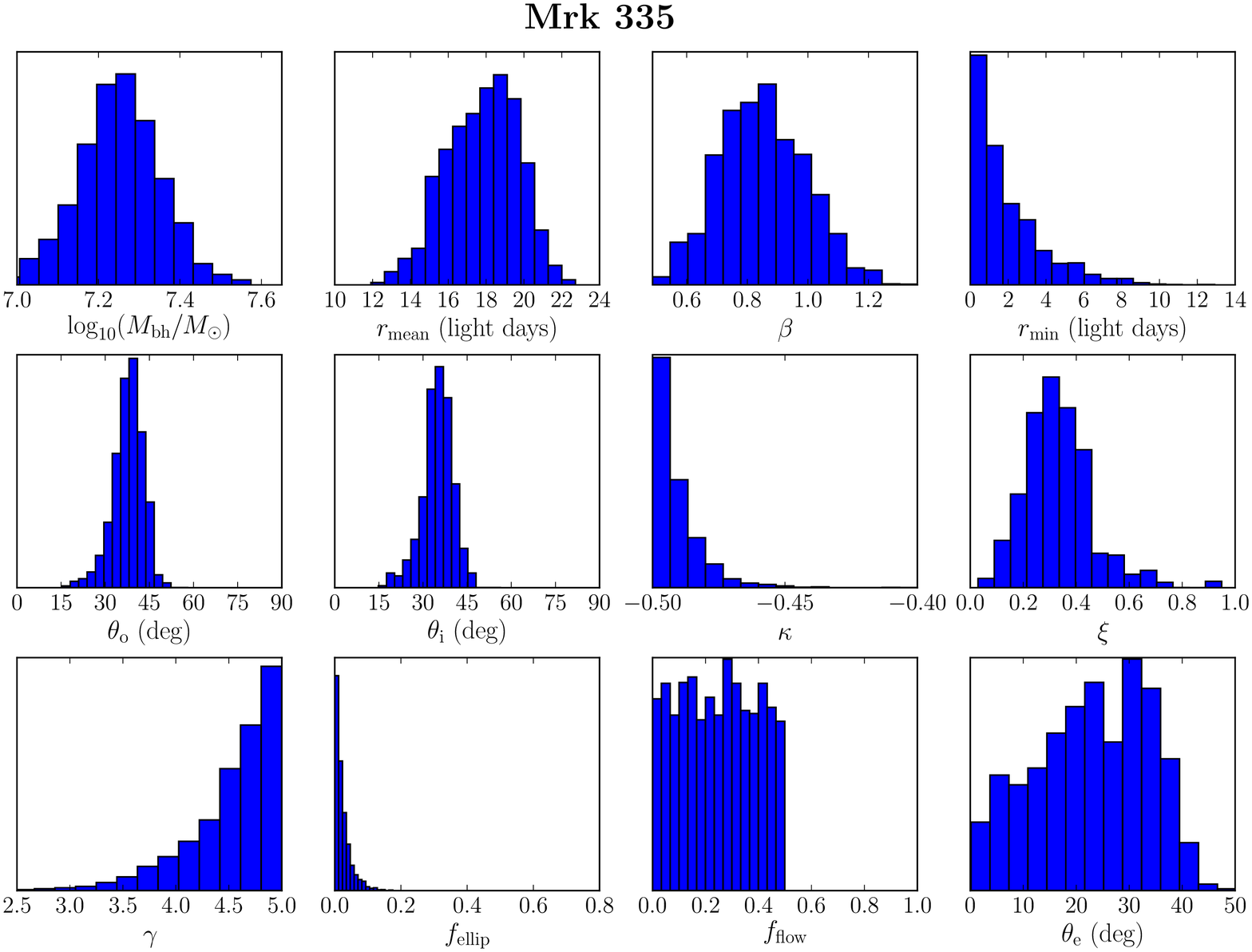}
\caption{Posterior distributions for key model parameters for Mrk\,335. }
\label{fig:mrk335_posteriors}
\end{center}
\end{figure*}

\begin{figure*}
\begin{center}
\includegraphics[scale = 0.45, angle = 0, trim = 0 0 0 0, clip]{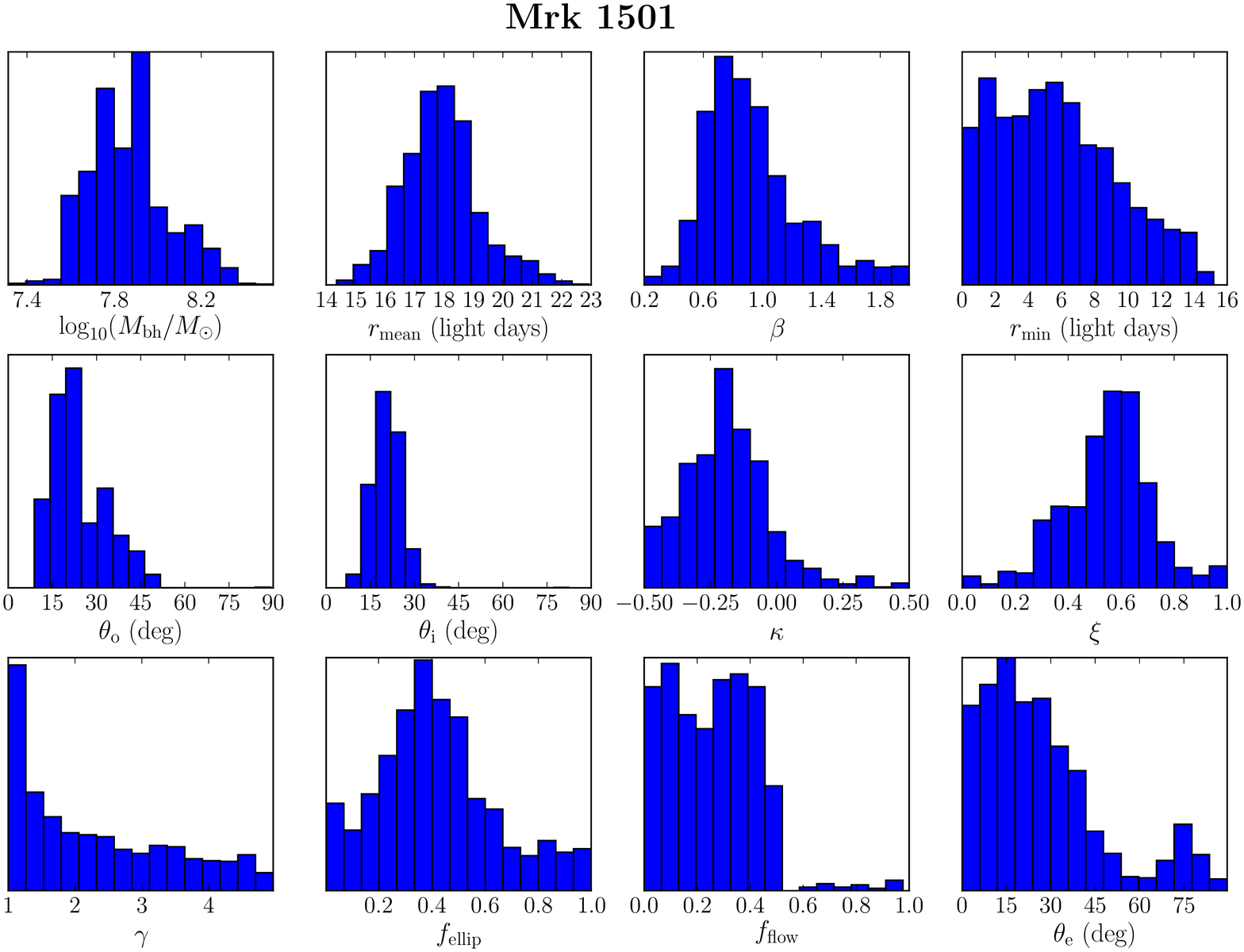}
\caption{Posterior distributions for key model parameters for Mrk\,1501. }
\label{fig:mrk1501_posteriors}
\end{center}
\end{figure*}

\begin{figure*}
\begin{center}
\includegraphics[scale = 0.45, angle = 0, trim = 0 0 0 0, clip]{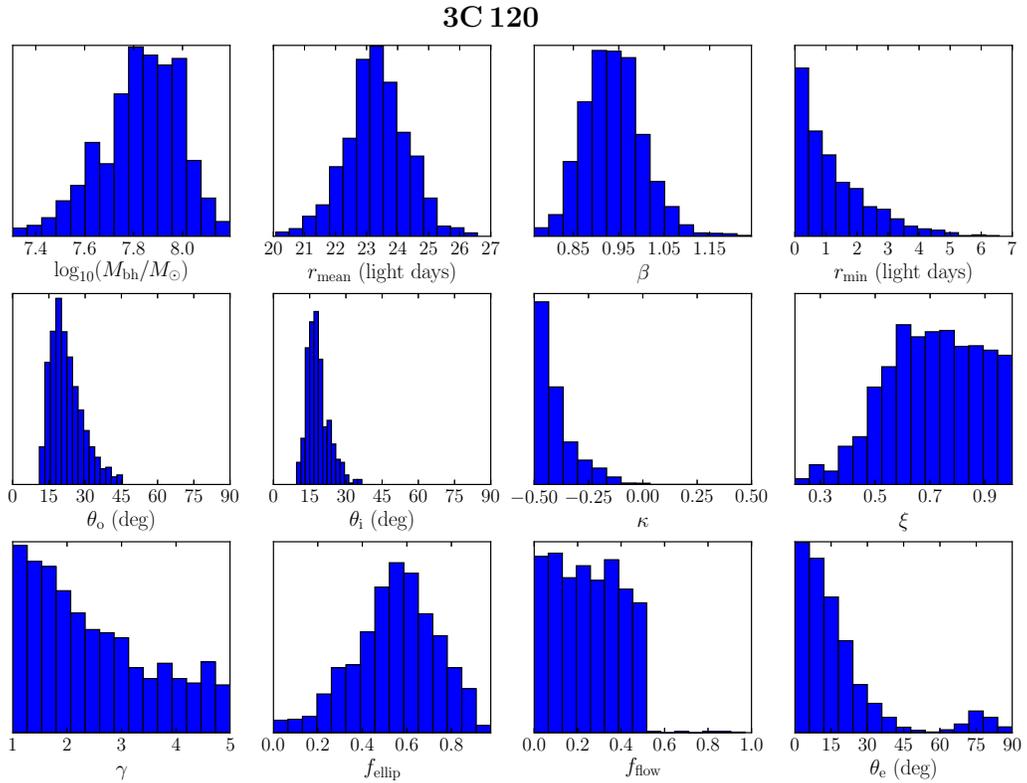}
\caption{Posterior distributions for key model parameters for 3C\,120.}
\label{fig:3c120_posteriors}
\end{center}
\end{figure*}

\begin{figure*}
\begin{center}
\includegraphics[scale = 0.45, angle = 0, trim = 0 0 0 0, clip]{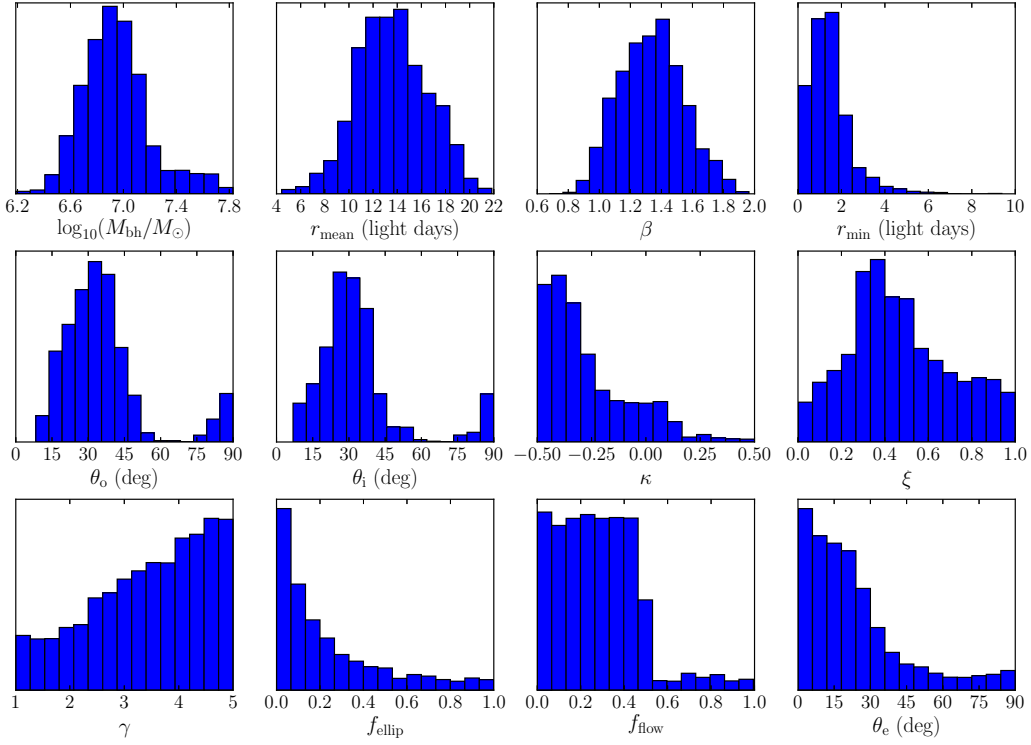}
\caption{Posterior distributions for key model parameters for PG\,2130+099. }
\label{fig:pg2130_posteriors}
\end{center}
\end{figure*}

\begin{figure*}
\begin{center}
\includegraphics[scale = 0.45, angle = 0, trim = 0 0 0 0, clip]{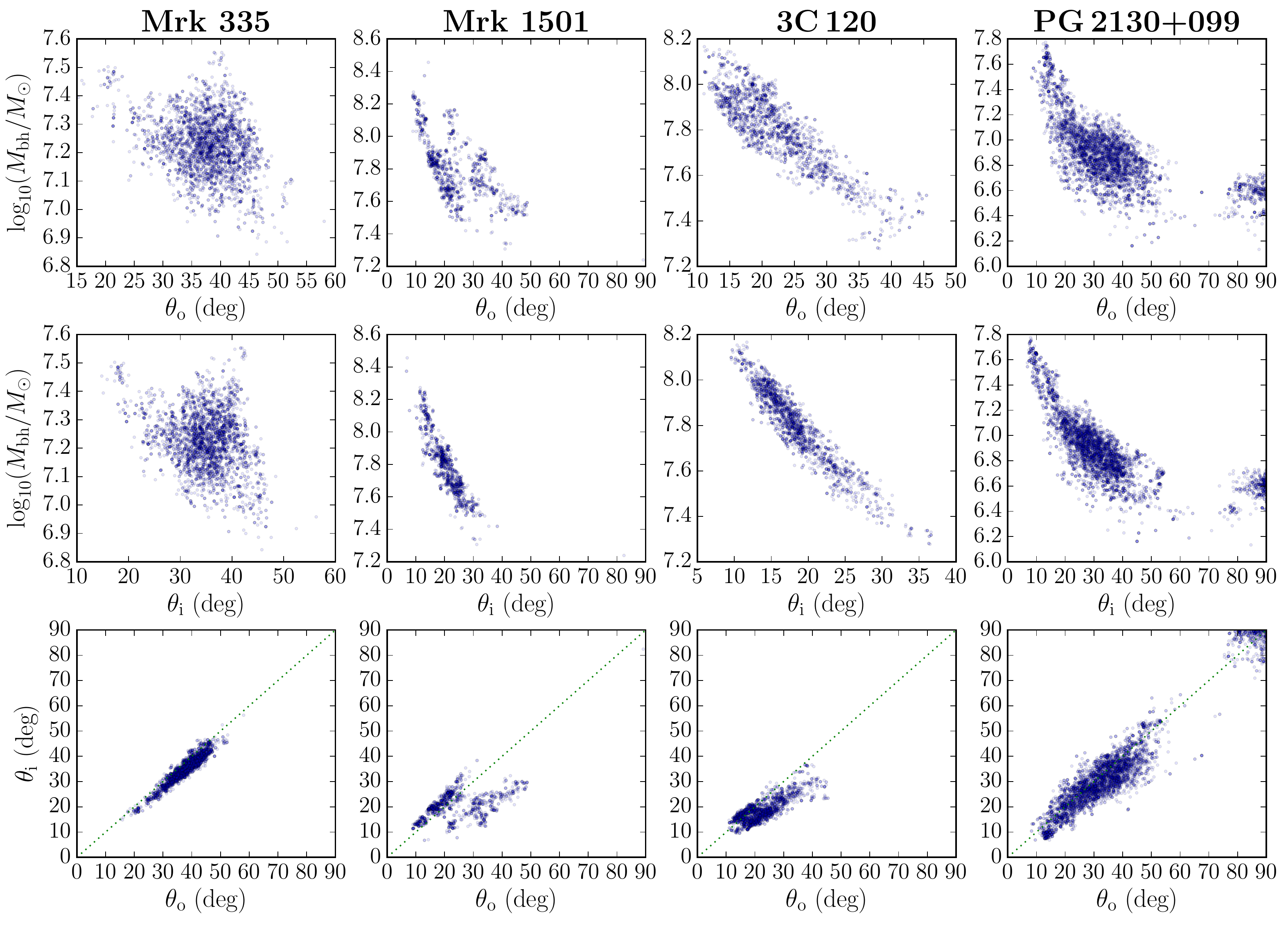}
\caption{Two-dimensional posterior distributions for all four AGN, showing the correlations between $M_{\rm BH}$, $\theta_i$, and $\theta_o$. }
\label{fig:2dposteriors}
\end{center}
\end{figure*}

\end{document}